\documentclass[runningheads,a4paper,draftcls]{llncs}

\usepackage{amssymb}
\usepackage{graphicx}
\usepackage[a4paper, total={6in, 8in}]{geometry}
\usepackage{times}
\usepackage[T1]{fontenc}
\usepackage{ae,aecompl}
\usepackage{amsmath}
\usepackage{fancyhdr}
\usepackage{multirow}
\usepackage{float}
\usepackage{subfig}
\usepackage{verbatim}
\usepackage{multirow}
\usepackage{xcolor}

\newcommand{\defi}{ \stackrel{\Delta}{=} }

\usepackage{url}
\urldef{\mailsa}\path|{rajinimakam,pruthvirajm,vssuresh}@iisc.ac.in|
\urldef{\mailsc}\path|sujit@iiserb.ac.in|    
\newcommand{\keywords}[1]{\par\addvspace\baselineskip
\noindent\keywordname\enspace\ignorespaces#1}

\begin{document}

\mainmatter  

\title{A Comprehensive Study on Modelling and Control of Autonomous Underwater Vehicle}

\titlerunning{Assistive Robotics}

\author{Rajini Makam$^1$ 
%
\thanks{(rajinimakam, pruthvirajm, vssuresh)@iisc.ac.in, sujit@iiserb.ac.in}%
\and Pruthviraj Mane$^2$ \and Suresh Sundaram$^3$ \and P.B. Sujit $^4$}
\institute{$^1$ 0000-0002-0989-9644, $^2$0009-0001-5047-4891,$^3$ 0000-0001-6275-0921, $^4$0000-0002-7297-1493}
\authorrunning{Assistive Robotics}


%
%

\toctitle{Assistive Robotics}
\tocauthor{Authors' Instructions}
\maketitle

\begin{abstract}
Autonomous underwater vehicles (AUV) have become the de facto vehicle for remote operations involving oceanography, inspection, and monitoring tasks. These vehicles operate in different and often challenging environments; hence, the design and development of the AUV involving hydrodynamics and control systems need to be designed in detail. This book chapter presents a study on the modelling and robust control of a research vehicle in the presence of uncertainties. The vehicle's dynamic behaviour is modelled using a 6-degree-of-freedom approach, considering the effect of ocean currents. The level flight requirements for different speeds are derived, and the resulting model is decomposed into horizontal and vertical subsystems for linear analysis.

The simulation results presented focus on the efficacy of linear controllers within three key subsystems: depth, yaw, and speed. Moreover, level-flight outcomes are demonstrated for a speed of 4 knots. The nonlinear control strategies employed in this study encompass conventional and sliding-mode control (SMC) methodologies. To ensure accurate tracking performance, the controller design considers the vehicle's dynamics with various uncertainties such as ocean currents, parameter uncertainty, CG (Center of Gravity) deviation and buoyancy variation. Both conventional and nonlinear SMC controllers' outcomes are showcased with a lawn-mowing manoeuvre scenario. A systematic comparison is drawn between the robustness of SMC against disturbances and parameter fluctuations in contrast to conventional controllers. Importantly, these results underscore the trade-off that accompanies SMC's robustness, as it necessitates a higher level of complexity in terms of controller design, intricate implementation intricacies, and the management of chattering phenomena.

\end{abstract}

\section{Introduction}

Vastness and inaccessibility of the ocean have led to several challenges in underwater exploration and research. There has been a growing interest in developing AUVs to achieve better exploration and data collection in the last few years. The AUVs are unmanned vehicles that are capable of operating on their own in underwater environments without direct human intervention. The development of these AUVs has opened several fields, especially in marine technology, marine biology, geology and oceanography. Because of their ability to navigate to difficult underwater terrains, and collect high-resolution scientific data, they have a key role in seafloor mapping, aircraft crash investigations, marine rescue, military applications and Intrusion surveillance. 

Depending on the specific application or operating condition there are different sizes, shapes and configurations of AUVs. Also, these applications require the AUV to carry different sensors like cameras, sonar, depth sensors, navigational sensors or some other scientific instrument. Although these developments and capabilities are fascinating, the underwater environment is characterized by its harsh conditions because of which there are several challenges that needs to be addressed. Due to the constraints of underwater wireless communication, AUVs have drawn more attention from the underwater scientific community. Autonomous decision-making underwater robots are becoming more dependable and beneficial due to improvements in AUV research, materials, manufacturing procedures, sensors, processing power, and battery technology. The goal of current AUV research is to develop a trustworthy, completely autonomous robotic decision-making system.

A "Fish"-like torpedo created by Robert Whitehead in 1886 is considered to be the first AUV in history \cite{Von2003}. 
The first torpedo had a range of 700 metres, was powered by compressed air, and had a top speed of 3.0 m/s. In 1957,  Special Purpose Underwater Research Vehicle (SPURV) was developed by the Applied Physics Laboratory at the University of Washington. The vehicle can dive to a depth of 3650 m and travel at a speed of 2-2.5 m/s \cite{Widditsch1973}. The goal was to collect oceanographic data using precise trajectory calculations. More than 400 deployments of SPURV were documented up until the late 1970s to carry out diverse duties before SPURV-II was built in 1973 \cite{Von2003}. 

In the last two decades, several AUVs have been developed by various organizations across the globe. Some noteworthy AUVs are namely Remote Environmental Monitoring UnitS (REMUS) and its variants \cite{Von2003}, \cite{ASA1997}, \cite{MBV2005} and \cite{JLG2019},  Autosub 6000 \cite{McPhail2009}, Indian origin AUVs MAYA \cite{BPS2004}, \cite{DMM2007}  and AUV 150 \cite{SNPD2012}, LAUV \cite{SMJ2012}, Odyssey AUV series \cite{CLV2005}, Bluefin12 and Bluefin21 \nocite{Bluefin, BluefinGD, FGPB2005} \cite{Bluefin} -\cite{FGPB2005}, HUGIN HUS AUV \cite{Hugin}. These AUVs are employed for surveillance, mine countermeasures, search and rescue, bathymetry, mapping the ocean floor, geological survey, environmental monitoring, repair and maintenance, structure inspection and several other underwater applications. Applications and technological challenges are tabulated in Table \ref{tapp}.

\begin{table}[]
\centering
\renewcommand{\arraystretch}{1.25}
\setlength{\tabcolsep}{7pt} 
\caption{AUV Applications}
\begin{tabular}{|c|c|c|}
\hline
                            & Application                                     & Technological challenges                                                                                 \\ \hline
\multirow{6}{*}{Military}   & Surveillance                                    &         \multirow{1}{*}{Localization,}      \\ \cline{2-2}
                            & Mine Countermeasures                            &  \multirow{1}{*}{path planning}     \\ \cline{2-2}
                            & Target localization                             &  \multirow{1}{*}{control,}               \\ \cline{2-2}
                            & Search and rescue                               &        \multirow{1}{*}{guidance,}                                             \\ \cline{2-2}
                            & Ocean exploration                               & \multirow{1}{*}{obstacle avoidance,}  \\ \cline{2-2}
                            & Anti-submarine warfare                         &    \multirow{1}{*}{trajectory tracking}                                                                                                      \\ \hline
\multirow{5}{*}{Scientific} & Geological survey                               &  \multirow{1}{*}{}                                 \\ \cline{2-2}
                            & Marine biology study                            &   \multirow{1}{*}{Obstacle avoidance, path planning,}\\ \cline{2-2}
                            & Archaeological survey                             &\multirow{1}{*}{trajectory tracking,}\\ \cline{2-2}
                            & Environmental monitoring                        &\multirow{1}{*}{control}\\ \cline{2-2}
                            & Seafloor mapping                               &   \multirow{1}{*}{}                                                                                                    \\ \hline
\multirow{2}{*}{Industry}   & Gas and oil inspection                        & \multirow{1}{*}{Localization, control, guidance}\\ \cline{2-2}
                            & Pipeline and infrastructure maintenance &\\ \hline
\multirow{3}{*}{Commercial} & Underwater video footage collection             & \multirow{3}{*}{Control, obstacle avoidance}                                               \\ \cline{2-2}
                            & Entertainment and Tourism                       &                                                                                                          \\ \cline{2-2}
                            & Fishing                                         &                                                                                                          \\ \hline
\end{tabular}
\label{tapp}
\end{table}

The successful operation of the AUVs depends on the mathematical model and  control architecture. They play a pivotal role in enhancing the vehicle capabilities and overall performance. The challenges AUVs face in underwater environments, such as complex hydrodynamics, uncertain external forces, and the need for precise manoeuvring, necessitates the development of sophisticated control strategies for stability, accuracy, and mission success. Over the past two decades, the research community has shown interest in employing neural networks for system identification and control, marking a significant advancement in the field. Pioneering work by Narendra \cite{NP1990} set the stage, followed by a sequence of endeavours that utilized actual vehicle data for system identification \nocite{KC1997, SOM2005, KSO2009, KOG2006, RC1999,   KCI2005, SK2008,  DSS2022}\cite{KC1997} - \cite{DSS2022}. Within the realm of adaptive neural network models, diverse architectures emerged, encompassing nonlinear neuroadaptive methods \cite{KC1997}, \cite{SOM2005}, \cite{KSO2009}, recurrent neural network \cite{KOG2006}, neural networks with augmented model inversion \cite{RC1999}, and the recent robust EMRAN approach for autonomous vehicles \cite{DSS2022}. While predominantly explored in aerial vehicles necessitating agile manoeuvres, these technologies are progressively making their mark in underwater vehicles. This transition extends to AUVs, where analogous identification methodologies have been presented for navigation \cite{CK2022}.  Further advancements in underwater vehicle speed control have leveraged self-propulsion estimation models \cite{KBR2020}, enabling the prediction of propeller rotation rates for a targeted speed. Notably, advancements in aggressive manoeuvres and redundant actuators, exemplified by technologies such as \cite{Hydrus}, are finding resonance in the field, poised to play a pivotal role in underwater vehicle systems. 

In spite of these advancements, most of these underwater vehicles are slow-moving and less agile in manoeuvring in nature. Moreover, one of the major challenges in the underwater environment is the disturbance due to the ocean current or the turbulence due to the motion of the large obstacles near the operation of the underwater vehicles. Hence most of the research focuses on conventional reliable robust control methods.
Over the years, numerous control architectures have been developed for AUVs that offer a wide array of controller options. Comprehensive survey articles and related references \nocite{SSF2017, SDS2019, KY2021, TCC2022, AXJ2023} \cite{SSF2017} - \cite{AXJ2023} provides an extensive overview of the diverse controllers used for AUVs. From the early stages of AUV experimentation, PID controllers emerged as a simple and popular choice \nocite{Brunner1988, HG1992, HM1992, Pres2001} \cite{Brunner1988} - \cite{Pres2001}. These controllers were employed for various AUV subsystems, including speed control, depth control, and steering control, often using linear PID designs \cite{herman2009}, \cite{KHS2011}, \cite{SDR2018}. For instance, the REMUS AUV utilizes a set point PD controller with regulator gain matrices to drive linear position and orientation errors to zero \cite{herman2009}. Further, fuzzy self-adapting PID controllers \cite{KB2015} and LMI-based PID controllers for wave disturbance suppression \cite{KFB2018} have been presented. In \cite{HGS2016}, an auto-tuned PID-like controller based on neural networks is designed. The PID controller's weights are tuned online to minimize position-tracking errors. To address parameter uncertainties, a fractional order PID controller has been designed for yaw control \cite{LZP2022}, offering robustness against variations of up to $+/-30\%$ from the actual parameters. 

Disturbance and perturbation are the main concerns in underwater vehicles. This affects the control performance of the system. A reliable method that is frequently applied to underwater vehicle control to provide robustness is Sliding Mode Control (SMC). This method provides robustness against bounded external perturbations and ensures finite-time convergence. Sliding mode control (SMC) has been widely used in literature \nocite{YS1984, HSC2010, JKY2014, TWTP2018} \cite{YS1984} - \cite{TWTP2018}. In early implementations, SMC was tested on a nonlinear underwater vehicle model using three different second-order controllers for position and velocity, demonstrating its effectiveness \cite{YS1984}. A modified SMC is implemented in \cite{CPH1990} to deal with changing dynamics and operating conditions in the dive plane. Authors in \cite{FF1995} present a robust adaptive sliding mode control law with a velocity state observer for a 3 DOF AUV model, highlighting its capability to handle disturbances and uncertainties in parameters. An SMC controller is designed for a 6-DoF model nonlinear system \cite{YYCM2013}. The robustness of SMC is demonstrated with the simulation results. Furthermore,  Decoupled systems, such as horizontal and vertical planes, are commonly used in AUVs with SMC, making them robust against bounded disturbances as shown in \cite{EZ2016}.

In \cite{CZC2016}, an adaptive SMC for pitch and yaw dynamics trajectory tracking is presented where the controller accounts for non-symmetric dead zones and unknown disturbances in the actuator. Additionally, an adaptive high-order sliding mode controller has been designed to track trajectories in the presence of persistent external disturbances \cite{GTC2019}. Complex AUV scenarios, such as highly coupled nonlinearities, unknown system parameters, disturbance uncertainties, and input saturation are dealt with in \cite{TTA2021}, \cite{DYW2022}. A method for handling the under-actuated AUV's trajectory tracking control with nonlinear model characteristics, model uncertainty, and roll motion effects is developed in \cite{TTA2021}.  Further, authors in \cite{DYW2022}, propose an adaptive backstepping sliding mode control method to mitigate uncertainties and disturbances, showcasing its superiority over conventional PID controllers in their comparative analysis. SMC has emerged as a versatile and robust control strategy for AUVs, enabling precise trajectory tracking and improved performance in challenging and uncertain underwater environments. Ongoing research in SMC and its combination with other control methodologies promises to further enhance AUV control capabilities and broaden their applications in various marine missions.

The focus of this book chapter is to provide a comprehensive study of modelling and control strategies for AUVs navigating in challenging underwater environments in the presence of ocean currents, parameters like buoyancy, hydro dynamic coefficients and centre of gravity variation. The foundation is laid through an intricate mathematical model that systematically integrates the impact of ocean currents into the AUV's dynamic behaviour, establishing a fundamental comprehension of its response within such complex surroundings. Further, the critical requirements and objectives for maintaining the desired level-flight of vehicles are investigated. During the level flight vehicle maintains constant velocity and orientation. This essentially involves finding one such control input that maintains the vehicle's equilibrium. It is seen in the literature the level-flight condition is largely ignored. Here, a detailed analysis of the level of flight dynamics is provided.  Subsequently, the dynamics in the horizontal plane, vertical plane and speed subsystems are derived. These subsystems aid in simplifying the control design process and facilitate stability analysis of the nonlinear system. Further, the simulation section provides insights into the performance of linear control strategies, conventional nonlinear controllers, and robust SMC. The nonlinear controllers are evaluated under varying conditions, including parameter variations, ocean currents, and centre of gravity fluctuations. Finally, an analysis of the robustness of two controllers: conventional and SMC is presented with the lawn mowing manoeuvre.  

The book chapter is organised as follows: in Section \ref{smod}, we lay the foundation by developing a comprehensive mathematical model that incorporates the influence of ocean currents on AUV dynamics. Subsequently, in Section \ref{strim}, we delve into the critical aspects of level flight dynamics and the linearization of the nonlinear model presented in Section \ref{smod}. In Section \ref{ssim}, we present the simulation results obtained through three distinct control approaches: linear control, nonlinear control using conventional controller, and SMC. Finally, Section \ref{scon} gives the concluding remarks.

\section{AUV Modelling}\label{smod}

\subsection{Research Vehicle Profile}
Getting the hydrodynamic coefficients that are required for the mathematical formulation of the vehicle analysis of the vehicle profile is essential. It includes hull, nose and tail section profiles, and locations of various points such as the centre of gravity, the centre of buoyancy and fin parameters.
The hull profile of the vehicle we are referring to is based on Myring hull profile \cite{Pres2001}. This profile is designed such that there is a minimal drag coefficient for the required length-to-diameter ratio. The sub-parameters required to define the profile are $a$ (length of nose section), $b$ (length of mid-hull section), $c$ (length of the tail section), $d$ (maximum diameter), $2\theta$ (included angle at the tip of tail) and $n$ (a parameter which when varied will give different body shapes).
Assuming an origin at the vehicle nose the equation of radius of curvature is given by

\begin{equation}
    r(\Xi)=\frac{1}{2} d\left[1-\left(\frac{\Xi+a_{\text {offset }}-a}{a}\right)^2\right]^{\frac{1}{n}},
\end{equation}
where, $r$ is the radius of the hull measured normally from the centerline, $\Xi$ is the axial position along the centerline and $a_{offset}$ is the missing vehicle nose length. Similarly, the tail section radius is given by 
\begin{equation}
r(\Xi)=\frac{1}{2} d-\left[\frac{3 d}{2 c^2}-\frac{\tan \theta}{c}\right](\Xi-l)^2+\left[\frac{d}{c^3}-\frac{\tan \theta}{c^2}\right]\left(\Xi-l_f\right)^3
\end{equation}
where $l$ is the forward body length and $l_f$ is given as 
\begin{equation}
l_f = a + b - a_{offset}.
\end{equation}
All these parameters are tabulated in Table \ref{tpara} in the appendix section. The general vehicle-related parameters along with the moment of inertia (MoI) parameters are tabulated in Table \ref{tmodpara}. MoI is with respect to the origin at the center of buoyancy (CB) which is located at 0.611 meters from the vehicle's nose. The location of the center of gravity (CG) wrt CB. The fin parameters and hull parameters are the same as in \cite{Pres2001}.
\begin{table}[]
\renewcommand{\arraystretch}{1.25}
\setlength{\tabcolsep}{7pt} 
\caption{Model Parameters of Research Vehicle}
\centering
\begin{tabular}{|cc|c|c|}
\hline
\multicolumn{2}{|c|}{Description}                                        & Unit     & Value  \\ \hline
\multicolumn{2}{|c|}{Vehicle mass ($m$)}                                 & $kg$     & 30.5   \\ \hline
\multicolumn{2}{|c|}{Seawater Density ($\rho$)}                          & $kg/m^3$ & 1030   \\ \hline
\multicolumn{2}{|c|}{Vehicle Buoyancy ($B$)}                             & $N$      & 306    \\ \hline
\multicolumn{2}{|c|}{MOI about x-axis ($I_{xx}$)}                        & $kg.m^2$ & 0.177  \\ \hline
\multicolumn{2}{|c|}{MOI about z-axis ($I_{zz}$)}                        & $kg.m^2$ & 3.45   \\ \hline
\multicolumn{1}{|c|}{\multirow{3}{*}{Center of Bouancy (CB)}} & $x_{cg}$ & $m$      & 0      \\ \cline{2-4} 
\multicolumn{1}{|c|}{}                                        & $y_{cg}$ & $m$      & 0      \\ \cline{2-4} 
\multicolumn{1}{|c|}{}                                        & $z_{cg}$ & $m$      & 0.0192 \\ \hline
\end{tabular} \label{tmodpara}
\end{table}

\subsection{Mathematical Model}
Mathematical modelling plays a fundamental role in the development and analysis of autonomous underwater vehicles (AUVs). AUVs are unmanned vehicles designed to operate autonomously in a challenging underwater environment, performing various tasks such as exploration, mapping, surveillance, and environmental monitoring. To ensure their efficient and safe operation, it is crucial to understand and predict the behaviour of these vehicles in different scenarios. A set of mathematical equations that capture the physical properties, kinematics, dynamics, and interactions with the environment is essential to gain insights into the behaviour and performance of AUVs. Accurate modelling of AUV dynamics and kinematics is fundamental to understanding their behaviour and interactions with the underwater environment. In summary, integrating accurate modelling and effective control strategies is indispensable for enhancing AUVs' autonomy, efficiency, and safety in underwater missions. 

The mathematical model of an AUV encompasses several important aspects. The dynamics of an AUV consist of two components: kinematics, which defines the vehicle's motion in terms of position, orientation, and velocity, and kinetics, which defines the 6- degrees of freedom (DoF) with the forces and torques acting on the AUV, accounting for hydrodynamic effects, buoyancy, thrust forces, and external disturbances \cite{Pres2001} and \cite{Fossen2011}. This dynamic representation allows for the study of the vehicle's response to different environmental conditions and control inputs.

\begin{figure}
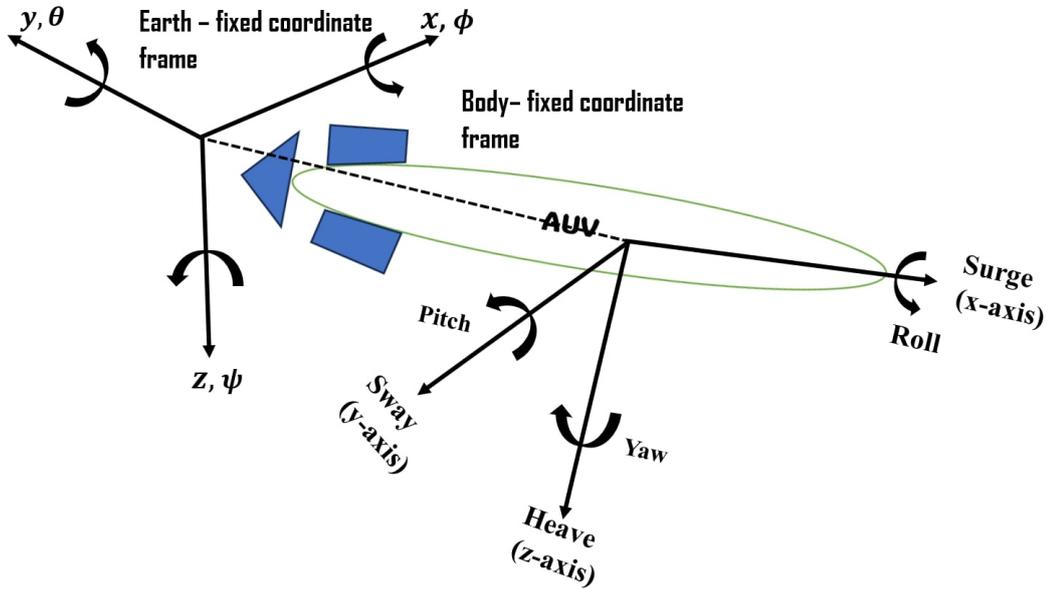

\centering
 \caption{Body-fixed and Earth-fixed Coordinate Systems} \label{coord}
\end{figure}

According to SNAME (Society of Naval Architects and Marine Engineers), AUV  is modelled with respect to two coordinate systems: body-fixed reference frame ${B}$ and earth-fixed reference frame ${G}$. These coordinate systems are shown in Fig \ref{coord}.  

Figure 1 cord.jpg here

The 6 DoF model of AUV has six independent coordinates expressed in a body-fixed frame that determines the motion of the vehicle. 
The first three coordinates represent linear velocities $\boldsymbol{\nu_1}$ while the rest three represent angular velocities $\boldsymbol{\nu_2}$.   
\begin{equation}
    \boldsymbol{\nu}=\left[\begin{array}{c}
        \boldsymbol{\nu_1} \\ \boldsymbol{\nu_2}
    \end{array}\right] =\left[ u ~ v~ w~ p~ q~ r\right]^T \label{eblvav}
\end{equation}
where $u$, $v$, and $w$ are linear velocities in the surge, sway, and heave axes 
respectively and $p$, $q$ and $r$ are angular velocities in roll, pitch and yaw respectively. The other coordinate system coincides with the earth's North-East-Down (NED) directions. The earth-fixed or inertial frame is fixed to a point on the water's surface. The position $\boldsymbol{\eta_1}$ and orientation $\boldsymbol{\eta_2}$ of the vehicle defined with respect to the earth-fixed frame respectively are,   
\begin{equation}
\boldsymbol{\eta}= \left[\begin{array}{c}
     \boldsymbol{\eta_1} \\ \boldsymbol{\eta_2}
\end{array}\right]= [x  \ y \ z \ \phi \ \theta \ \psi]^T \label{eepo}
\end{equation}
Lastly, total forces $\boldsymbol{\tau_1}$ and moments $\boldsymbol{\tau_2}$ acting on the vehicle are expressed in the body-fixed frame as  \\
\begin{equation}
    \boldsymbol{\tau} =\left[\begin{array}{c}
          \boldsymbol{\tau_1} \\ \boldsymbol{\tau_2} 
    \end{array}\right]=\left[X\ Y\ Z\ K\ M\ N\right]^T
\end{equation}
where $X$, $Y$ and $Z$ represents forces and $K$, $M$ and $N$ represents moments in $B$. 
The following coordinate transform relates the time derivatives of $\boldsymbol{\eta}$ to $\boldsymbol{\nu}$: 
\begin{equation} 
\left[\begin{array}{c}\boldsymbol{\dot{\eta_1}(t)}\\\boldsymbol{\dot{\eta_2}(t)} \end{array}\right]_G=\left[\begin{array}{cc}
    \boldsymbol{T^G_B(\eta_2)}  & 0_{3x3} \\
     0_{3x3}       & \boldsymbol{A^G_B(\eta_2)}\end{array}\right] \label{ekin}
\left[\begin{array}{c}\boldsymbol{\dot{\nu_1}(t)}\\\boldsymbol{\dot{\nu_2}}\left(t\right) \end{array}\right]_B
\end{equation}
where $0_{3\text{x}3}$ is a null matrix of dimension 3, $\boldsymbol{T^G_B}$ transforms the linear velocities
\begin{equation}\boldsymbol{T^G_B\left(\eta_2\right)}=\ \left[\begin{array}{ccc}\cos{\psi}\cos{\theta}&-\sin{\psi}\cos{\phi}+\cos{\psi}\sin{\theta}\sin{\phi}&\sin{\psi}\sin{\phi}+\cos{\psi\sin{\theta\cos{\phi}}}\\\sin{\psi\cos{\theta}}&\cos{\psi\cos{\phi+\sin{\psi\sin{\theta\sin{\phi}}}}}&-\cos{\psi\sin{\phi+\sin{\psi\sin{\theta\cos{\phi}}}}}\\-\sin{\theta}&\cos{\theta\sin{\phi}}&\cos{\theta\cos{\phi}}\\\end{array}\right]
\end{equation}
and the transformation matrix $\boldsymbol{T^G_B(\eta_2)}$ satisfies orthogonality property,
$$ \boldsymbol{T^G_B}\left(\boldsymbol{\eta_2}\right)^{-1}=\boldsymbol{T^G_B}\left(\boldsymbol{\eta_2}\right)^T $$
and $\boldsymbol{A^G_B}$ transformation matrix between time derivative of Euler angles  and angular velocities and given by,
\begin{equation}
\boldsymbol{A^G_B}\left(\boldsymbol{\eta_2}\right) = 
\left[\begin{array}{ccc}1&\sin{\phi}\tan{\theta}&\cos{\phi\tan{\theta}}\\0&\cos{\phi}&-\sin{\phi}\\0&\frac{\sin{\phi}}{\cos{\theta}}&\frac{\cos{\phi}}{\cos{\theta}}\\\end{array}\right] 
\end{equation}

The equations of motion can be derived using the Newton–Euler or Lagrange equations. In order to derive the equations of motion, it is necessary to study the motion of rigid bodies, hydro-statics and hydrodynamics. The system can be represented in the form of 6 DOF equations defined in body-fixed coordinates: wherein the three force equations in $u,~v, ~w$ axes are: 
\begin{equation}
\begin{split}
 m[\dot{u}\ -vr+wq\ -x_g\left(q^2-r^2\right)+y_g\left(pq-r^2\right)+z_g\left(pr+\dot{q}\right) = X \\ \label{e6dof1}
 m[\dot{v}\ -wp+ur\ -y_g\left(p^2+r^2\right)+z_g\left(qr-\dot{p}\right)+x_g\left(pq+\dot{r}\right) = Y \\
 m[\dot{w}\ -uq+vp\ -z_g\left(p^2+q^2\right)+x_g\left(rp-\dot{q}\right)+y_g\left(rq+\dot{p}\right) = Z  
\end{split}
\end{equation}
Similarly, moment balance in three axes leads to the following moment equations: 
\begin{equation}
\begin{split} 
& I_{xx}\ \dot{p}+\left(I_{zz}-I_{yy}\right)qr+m\left[y_g\left(\dot{w}-uq+vp\right)-z_g\left(\dot{v}-wp+ur\ \right)\right] = K \\ \label{e6dof2}
& I_{yy}\ \dot{q}+\left(I_{xx}-I_{zz}\right)rp+m\left[z_g\left(\dot{u}-vr+wq\right)-x_g\left(\dot{w}-uq+vp\right)\right] = M \\
& I_{zz}\ \dot{r}+\left(I_{yy}-I_{xx}\right)pq+m\left[x_g\left(\dot{v}-wp+ur\right)-y_g\left(\dot{u}-vr+wq\right)\right] = N 
\end{split}
\end{equation}
where 
$m$ is the mass of the AUV, $x_g, y_g, z_g$ forms the location of
center of gravity ($C_g$), $I_{xx},~I_{yy},~I_{zz}$ are moments of 
the inertia of AUV, and the external forces :
\begin{equation}
\begin{split} 
  X   = &  {X}_{ {HS}}+ {X}_{ {u}\left| {u}\right|} {u}\left| {u}\right|+ {X}_{\dot{ {u}}}\dot{ {u}}+ {X}_{ {wq}} {wq}+ {X}_{ {q|q|}} {q|q|}+ {X}_{ {vr}} {vr}+ {X}_{ {rr}} {rr}+ {X}_{ {prop}}  \\
 Y = &
 {Y}_{ {HS}}+ {Y}_{ {v}\left| {v}\right|} {v}\left| {v}\right|+ {Y}_{ {r}\left| {r}\right|} {r}\left| {r}\right|+ {Y}_{\dot{ {v}}}\dot{ {v}}+ {Y}_{\dot{ {r}}}\dot{ {r}}+ {Y}_{ {ur}} {ur}+ {Y}_{ {wp}} {wp}+ {Y}_{ {pq}} {pq}+ {Y}_{ {uv}} {uv} \\& 
 + {Y}_{ {uu} {\delta}_ {r}} {u}^2(\delta_{r_1} + {\delta}_ {r_2})\\ %
 Z  = &  {Z}_{{HS}}+{Z}_{{w}\left|{w}\right|}{w}\left|{w}\right|+{Z}_{{q}\left|{q}\right|}{q}\left|{q}\right|+{Z}_{\dot{{w}}}\dot{{w}}+{Z}_{\dot{{q}}}\dot{{q}}+{Z}_{{uq}}{uq}+{Z}_{{vp}}{vp}+{Z}_{{rp}}{rp}+{Z}_{{uw}}{uw} \\&
 +{Z}_{{uu}{\delta}_{s}}{u}^2(\delta_{s_1} + {\delta}_{s_2}) \label{force_equations} 
\end{split}
 \end{equation}
 An external moments $K,~M,~N$ are given by:
 \begin{equation}
\begin{split} 
K = &  {K}_{ {HS}}+ {K}_{ {p}\left| {p}\right|} {p}\left| {p}\right|+ {K}_{\dot{ {p}}\ }\dot{ {p}}+ {K}_{ {prop}} + {K_{roll}\delta_{roll}} \\ 
 M = &  {M}_{ {HS}}+ {M}_{ {w}\left| {w}\right|} {w}\left| {w}\right|+ {M}_{ {q}\left| {q}\right|} {q}\left| {q}\right|+ {M}_{\dot{ {w}}}\dot{ {w}}+ {M}_{\dot{ {q}}}\dot{ {q}}+ {M}_{ {uq}} {uq}+ {M}_{ {vp}} {vp}+ {M}_{ {rp}} {rp}+ {M}_{ {uw}} {uw} \\ &
 + {M}_{ {uu} {\delta}_ {s}} {u}^2(\delta_{s_1} + {\delta}_{s_2})\\
 N = & {N}_{ {HS}}+ {N}_{ {v}\left| {v}\right|} {v}\left| {v}\right|+ {N}_{ {r}\left| {r}\right|} {r}\left| {r}\right|+ {N}_{\dot{ {v}}}\dot{ {v}}+ {N}_{\dot{ {r}}}\dot{ {r}}+ {N}_{ {ur}} {ur}+ {N}_{ {wp}} {wp}+ {N}_{ {pq}} {pq}+ {N}_{ {uv}} {uv} \\& + {N}_{ {uu} {\delta}_ {r}} {u}^2 (\delta_{r_1} + {\delta}_ {r_2})\label{moment_equations} 
\end{split}
 \end{equation}
All the coefficients in eq. \eqref{force_equations} and eq. \eqref{moment_equations} are given in Table \ref{tfc}. The matrix representation of eq. \eqref{e6dof1} and eq. \eqref{e6dof2},
\begin{eqnarray}
    \left[\begin{array}{c}
         \dot{u}  \\  \dot{v}  \\ \dot{w}  \\ \dot{p}  \\ \dot{q}  \\ \dot{r} 
    \end{array}\right] = 
    \left[\begin{array}{cccccc}
         m-X_{\dot{u}} & 0 & 0 & 0 & mz_g & -my_g  \\  
         0 & m-Y_{\dot{v}} & 0 & -mz_g & 0 & mx_g - Y_{\dot{r}} \\ 
         0 & 0 & m-Z_{\dot{w}} & my_g & -mx_g-Z_{\dot{q}} & 0 \\
         0 & -mz_g & my_g & I_{xx} - K_{\dot{p}} & 0 & 0  \\ 
         mz_g & 0 & -mx_g - M_{\dot{w}} & 0 & I_{yy}-M_{\dot{q}} & 0 \\ 
         -my_g & mx_g - N_{\dot{v}} & 0 & 0 & 0 & I_{zz}-N_{\dot{r}}
    \end{array}\right]^{-1} 
    \left[\begin{array}{c}
         X  \\ Y  \\ Z  \\ K  \\ M  \\ N 
    \end{array}\right] \label{ematrix}
\end{eqnarray}
While eq. \eqref{ematrix} offers a concise representation of the AUV's nonlinear model, its applicability may be limited when accounting for factors like ocean currents or external disturbances. The utilization of a modular system representation, on the other hand, presents distinct advantages during controller design, enabling the development of strategies specifically tailored to counteract the influence of external perturbations. In general, eq. \eqref{ematrix} could be expressed in the following standard form, 
\begin{equation}
     \boldsymbol{M\dot{\nu} +  C(\nu) \nu + D(\nu) \nu + g(\eta) = \tau}  \label{evect} 
\end{equation}
   where $\boldsymbol{M}$ is the mass and inertial matrix, $\boldsymbol{C(\nu)}$ is Coriolis and centripetal matrix, $\boldsymbol{D(\nu)}$ is damping matrix, $\boldsymbol{g(\eta)}$ is the hydrostatic vector and $\boldsymbol{\tau}$ is the control input vector. The details of the individual terms in eq. \eqref{evect} are given below;
\begin{enumerate}
    \item {\bf Matrix $\boldsymbol{M}$}: It is composed of rigid-body system inertia matrix $\boldsymbol{M_{RB}}$ and added mass matrix $\boldsymbol{M_A}$,
\begin{equation}
    \boldsymbol{M = M_{RB} + M_A} \label{eM}
\end{equation}
The rigid-body system inertia matrix $M_{RB}$
\begin{equation}
 \boldsymbol{M_{RB}} = \left[\begin{array}{cccccc}
       m & 0 & 0 & 0 & mz_g & -my_g \\
       0 & m & 0 & -mz_g & 0 & mx_g \\
       0 & 0 & m & my_g & -mx_g & 0 \\ \label{eMRB}
       0 & -mz_g & my_g & I_{xx} & 0 & 0 \\
       mz_g & 0 & -mx_g & 0 & I_{yy} & 0 \\
       -my_g & mx_g & 0 & 0 & 0 & I_{zz}
    \end{array}\right] 
\end{equation}
The amount of water that is added to the vehicle's mass as it accelerates is known as added mass. When a vehicle accelerates or decelerates, the volume of the surrounding fluid must be changed, which results in a pressure-induced force and moment on the body. This effect is substantial for objects moving through the water, and its contribution can be described as an apparent rise in the real mass of the submerged object. The added mass matrix $M_A$ expressed as,
\begin{equation}
  \boldsymbol{M_A} = - \left[\begin{array}{cccccc}
       X_{\dot{u}} & 0 & 0 & 0 & 0 & 0 \\ \label{eMA}
       0 & Y_{\dot{v}} & 0 & 0 & 0 & Y_{\dot{r}} \\
       0 & 0 & Z_{\dot{w}} & 0 & Z_{\dot{q}} & 0 \\
       0 & 0 & 0 & K_{\dot{p}} & 0 & 0 \\
       0 & 0 & M_{\dot{w}} & 0 & M_{\dot{q}} & 0 \\
       0 & N_{\dot{v}} & 0 & 0 & 0 & N_{\dot{r}}
    \end{array}\right] 
\end{equation}
where $X_{\dot{u}},~ Y_{\dot{v}},~Y_{\dot{r}}, ~Z_{\dot{w}},~Z_{\dot{q}},~ K_{\dot{p}},~M_{\dot{w}},~M_{\dot{q}},~N_{\dot{v}},~N_{\dot{r}}$ are hydrodynamic added mass coefficients
Calculation of these coefficients for a generic shape is a challenging problem that requires the use of CFD techniques and experimental verification. However, there are a number of streamlined techniques that offer suitable first-order approximations for frequently seen shapes (such as spheres, cylinders, plates, etc.) \cite{Pres2001}.\\

\item {\bf Coriolis and Centripetal matrix $\boldsymbol{C(\nu)}$}:
This matrix is composed of two entities, the rigid body Coriolis and centripetal matrix $\boldsymbol{C_{RB}(\nu)}$ and Coriolis and centripetal effect due to added mass $\boldsymbol{C_A(\nu)}$. 
\begin{equation}
\boldsymbol{C(\nu) = C_{RB}(\nu) + C_A(\nu)} \label{ecormat}
\end{equation}
The matrix captures the effects of the Coriolis and centripetal force on the motion of the AUV. The Coriolis force arises due to the vehicle's motion in a rotating frame of reference, such as the Earth. While centripetal forces arise due to the vehicle's curved motion or manoeuvres. As the AUV moves through the water, its motion experiences a force that tends to deflect its trajectory sideways. In general, the matrix $C_{RB}$ in eq. \eqref{ecormat} is given by,
\begin{equation}
   \boldsymbol{C_{RB}(\nu)} = \left[\begin{array}{ccc}
            0 & 0 & 0  \\
            0 & 0 & 0  \\
            0 & 0 & 0 \\
            -m(y_gq + z_gr) & m(y_gp + w) & m(z_gp - v) \\ 
             m(x_gq - w) & -m(z_gr + x_gp) & m(z_gq + u)  \\
            m(x_gr + v)  & m(y_gr - u) & -m(x_gp + y_gq) \end{array}\right. \end{equation}
          \[ \left. \begin{array}{ccc}
              m(y_gq + z_gr) & -m(x_gq - w) & -m(x_gr + v) \\
             -m(y_gp + w) & m(z_gr + x_gp) & -m(y_gr - u) \\
             -m(z_gp - v) & -m(z_gq + u) & m(x_gp + y_gq)  \\
             0 & -I_{yz}q- I_{xz}p + I_{zz}r & I_{yz}r + I_{xy}p - I_{yy}q \\
             I_{yz}q + I_{xz}p -I_{zz}r & 0 & -I_{xz}r-I_{xy}q+I_{xx}p \\
             -I_{yz}r - I_{xy}p + I_{yy}q & I_{xz}r + I_{xy}q - I_{xx}p & 0 
            \end{array}\right] \]
Due to the fact that the vehicle has two axial planes of symmetry and homogeneous mass distribution, $I_{xy}$, $ I_{xz}$ and $I_{yz}$ are negligible compared to $I_{xx}$, $I_{yy} $ and $I_{zz}$. They are assumed to be zeros. Assuming that the $B$-frame coordinate origin is set in the centre line of the AUV, hence, $y_g = 0$.
The matrix $\boldsymbol{C_A}$ in eq. \eqref{ecormat} is analogous to $\boldsymbol{M_A}$ in eq. \eqref{eMA} which takes in the additional effects due to the motion of the vehicle in the fluid. 
\begin{equation} 
\boldsymbol{C_A(\nu)} = \left[\begin{array}{cccccc} 0 & 0 & 0  & 0 & -Z_{\dot{w}}w & Y_{\dot{v}}v \\
                    0 & 0 & 0 & Z_{\dot{w}}w  & 0 & -X_{\dot{u}}u \\
                    0 & 0 & 0 & -Y_{\dot{v}}v  & X_{\dot{u}}u & 0 \\
                    0 & -Z_{\dot{w}}w & Y_{\dot{v}}v & 0 & -N_{\dot{r}}r & M_{\dot{q}}q \\
                    Z_{\dot{w}}w & 0 & -X_{\dot{u}}u  & N_{\dot{r}}r & 0 & 
                    -K_{\dot{p}}p \\
                    -Y_{\dot{v}}v & X_{\dot{u}}u & 0 & -M_{\dot{q}}q & K_{\dot{p}}p & 0 \end{array}\right]
\end{equation}
\\

\item {\bf Hydrodynamic Damping matrix $\boldsymbol{D(\nu)}$}: Hydrodynamic damping is a critical aspect of the hydrodynamic forces acting on an AUV. It is usually written as the sum of linear and non-linear damping,
\begin{equation} 
\boldsymbol{D(\nu) = D_L + D_n(\nu)} \label{edampmat}
\end{equation}
The linear damping matrix $D_L$ is due to potential damping and skin friction and the nonlinear damping matrix $\boldsymbol{D_n(\nu)}$ is due to quadratic damping and higher-order terms.  As mentioned it arises due to various factors, including skin friction, vortex shedding, and lifting forces. Skin friction is the resistance to motion experienced by the AUV as it interacts with the water. When the AUV's hull moves through the water, the friction between the hull's surface and the water generates drag forces. Skin friction damping is influenced by factors such as the hull's shape, the roughness of the surface, and the vehicle's speed. An AUV with a propeller moves through the water, it can generate vortices in its wake due to the flow separation around the vehicle's body. These vortices can lead to vortex shedding, a phenomenon where vortices are periodically shed from the AUV's surface. Vortex shedding introduces additional drag forces on the AUV. Lifting forces, also known as hydrodynamic lift, occur when there is a pressure difference between the upper and lower surfaces of the AUV's hull. 
The linear damping matrix $\boldsymbol{D_L}$ in eq. \eqref{edampmat} can be written as,
\[\boldsymbol{D_L} = - \left[\begin{array}{cccccc}
                X_u & 0 & 0 & 0 & 0 & 0 \\
                0 & Y_v & 0 & Y_p & 0 & Y_r \\
                0 & 0 & Z_w & 0 & Z_q & 0 \\
                0 & K_v & 0 & K_p & 0 & K_r \\
                0 & 0 & M_w & 0 & M_q & 0 \\
                0 & N_v & 0 & N_p & 0 & N_r 
             \end{array} \right]\]
Certain cross-terms can be neglected when AUV is operating near the equilibrium point.
The nonlinear damping matrix $\boldsymbol{D_n(\nu)}$ has two components,
\begin{equation}
\boldsymbol{D_n(\nu) = D_{n_1}(\nu) + D_{n_2}(\nu)} \label{eDnl}
\end{equation}
where, $\boldsymbol{D_{n_1}(\nu)}$ is the expression of crossflow drag terms,
\begin{equation}
    \boldsymbol{D_{n_1}(\nu)} = - \left[\begin{array}{cccccc} 
                    X_{|u|u}|u|  & 0 & 0 & 0 & 0 & 0 \\
                    0 & Y_{|v|v}|v| & 0 & 0 & 0 & Y_{|r|r}|r| \\
                    0 & 0 & Z_{|w|w}|w| & 0 & Z_{|q|q}|q| & 0 \\
                    0 & 0 & 0 & K_{|p|p}|p| & 0 & 0 \\
                    0 & 0 & M_{|w|w}|w| & 0 & M_{|q|q}|q| & 0 \\
                    0 & N_{|v|v}|v| & 0 & 0 & 0 & N_{|r|r}|r|
                    \end{array} \right]
\end{equation}
and  $\boldsymbol{D_{n_2}(\nu)}$ deals with the lift force and moments on the hull and fins of AUV,
\begin{equation}
    \boldsymbol{D_{n_2}(\nu)} = - \left[\begin{array}{cccccc} 
                    0  & 0 & 0 & 0 & 0 & 0 \\
                    0 & Y_{uv}u & 0 & 0 & 0 & Y_{ur}u \\
                    0 & 0 & Z_{uw}u & 0 & Z_{uq}u & 0 \\
                    0 & 0 & 0 & K_{up}u & 0 & 0 \\
                    0 & 0 & M_{uw}u & 0 & M_{uq}u & 0 \\
                    0 & N_{uv}u & 0 & 0 & 0 & N_{ur}u
                    \end{array} \right]
\end{equation}
           
\item {\bf Hydrostatic/ Resorting force vector} {$\boldsymbol{g(\eta)}$}: The vehicle experiences hydrostatic forces and moments as a result of the combined effects of the vehicle weight and buoyancy. The vehicle weight is given by $ W = mg $ and buoyancy is expressed as $ B = \rho V_mg$, where $\rho$ is the density of the surrounding fluid and $V_m$ is the total volume displaced by the vehicle. These forces and moments are expressed in
$B$. The gravitational and buoyancy forces acting on the centre of gravity $\boldsymbol{(c_g)}$ and centre of buoyancy $\boldsymbol{(c_b)}$ of an AUV are shown in Fig.\ref{fighs}

\begin{figure}
\centering
 \includegraphics[scale=0.6]{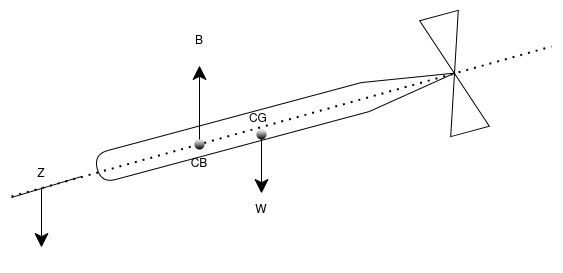} 
 \caption{Hydrostatic forces acting on AUV \cite{HA2011}} \label{fighs}
\end{figure}
\begin{equation} \label{ehs}
g(\eta) = \left[ \begin{array}{c}
     X_{HS}  \\ Y_{HS} \\ Z_{HS} \\ K_{HS} \\ M_{HS} \\ N_{HS}     
\end{array}\right] = 
      \left[ \begin{array}{c}-\left(W-B\right)\sin{\theta} \\
      \left(W-B\right)\cos{\theta\sin{\phi}} \\
      \left(W-B\right)\cos{\theta\cos{\phi}} \\
      -\left(y_gW-y_bB\right)\cos{\theta\cos{\phi}\ -\left(z_gW-z_bB\right)\cos{\theta\sin{\phi}}}\\
       -\left(z_gW-z_bB\right)\sin{\theta\ -\left(x_gW-x_bB\right)\cos{\theta\cos{\phi}}}\\
       -\left(x_gW-x_bB\right)\cos{\theta\sin{\phi}\ }-\left(y_gW-y_bB\right)\sin{\theta} \end{array}\right]
\end{equation}
where, $\boldsymbol{c_b} = [x_b~ y_b~ z_b]^T$ are vehicle's centers of buoyancy. \\

\item {\bf Control Input $\boldsymbol{\tau}$}: It consists of the terms related to the thrust and torque of the propeller and control surfaces: sterns $\delta_{s_1}$ and $\delta_{s_2}$   and rudders $\delta_{r_1}$ and $\delta_{r_2}$.  
\begin{equation}
    \boldsymbol{\tau} = \left[\begin{array}{c}
     X_{prop} \\  Y_{r} \\  Z_{s} \\ K_{prop} \\ M_s \\  N_r \end{array}\right] 
     = \left[\begin{array}{c} \left(1-\tau_p\right) T_{n|n|} n|n| \\ Y_{uu{\delta_r}}u^2 (\delta_{r_1} + \delta_{r_2}) \\   Z_{uu{\delta_s}}u^2 (\delta_{s_1} + \delta_{s_2}) \\ Q_{n|n|}n|n| + K_{roll}\delta_{roll} \\ M_{uu{\delta_s}}u^2 (\delta_{s_1} + \delta_{s_2}) \\ N_{uu{\delta_r}}u^2 (\delta_{r_1} + \delta_{r_2})     
     \end{array}\right]
\end{equation}
where $Y_{uu{\delta_r}}, Z_{uu{\delta_s}}, m_{uu{\delta_s}}, N_{uu{\delta_r}}$ are fin lift coefficients and $X_{prop} $ and $K_{prop}$ are the thrust and the torque of the propeller system. The thrust and torque are obtained from the model basin or thrust bench. Axial inflow ${u_p}$ is the main reason for thrust losses. This is generally neglected while designing the speed controller, eventually leading to thrust degradation. To account for this \cite{HRCMB1994} presented a model that justified the anomalies between experimental and mathematical thrust data. The two-state model is: 

\begin{equation}
\begin{split}
{J_m}\dot{n}  & =  Q - {K_{n}n} - {Q_{n|n|}n|n|}  \\
{m_f}\dot{u_p} & =  {T_{n|n|}n|n|} - {{d_f}{u_p}} - {{d_f}|u_p|({u_p}-(1-{w_p})u)} \label{ethrust}
\end{split}
\end{equation}
\\
where $d_{f0}$ and $d_f$ are damping coefficients given by, 
\begin{equation}
\begin{split}
    d_{f0} = \frac{-2X_{u|u|}}{(1-{\tau_p})(1+{a_p})(1-{w_p})}, \\ \label{edampcoff}
d_f = \frac{-X_{u|u|}}{(1-{\tau}_p)(1+a_p)(1-w_p)^2},
\end{split}
\end{equation}
where $Q$ is propeller torque, $J_m$ is propeller inertia, $m_f$ if mass in control volume, $T_{n|n|}$ and $Q_{n|n|}$ are thrust and torque coefficients respectively, $\tau_p$ is thrust reduction factor, $a_p$ is axial flow parameter, $w_p$ is thrust wake fraction number. Another simple model for the propulsion system, treats the propeller as a source of constant thrust and torque. The experimental coefficient data obtained by \cite{Pres2001} are used for mathematical simulation. Based on the experimental data at different speeds, the thrust and torque can be calculated using a second-order polynomial for the given speed of the propeller of the AUV.\\ Propeller force on AUV in X-direction as a function of rotational speed can be approximated as,
 \begin{equation} \label{exprop}
  X_{prop} =  a_1 * n^2 + a_2*n + a_3
    \end{equation}
and similarly, the torque along the X-direction as a result of the propeller rotation be, 
   \begin{equation}
    K_{prop} =  b_1 * n^2 + b_2*n + b_3 \label{ekprop}
    \end{equation}
where $n$ is the speed of the propeller in rotations per second (RPS) and $(a_1, a_2, a_3)$ and $(b_1, b_2, b_3)$ are constants. To simplify the equations we use second-order polynomial approximation as the second-order curve fitting for thrust vs. speed gives an accuracy of $98.32\%$.
\end{enumerate}
\subsubsection{Ocean currents:}
The nonlinear state-space model in eq. \eqref{evect} has no effect on external disturbances. However, in practical scenarios, the system may be subjected to temporary or persistent ocean currents. Ocean currents are driven by gravity, wind (Coriolis Effect) and water density. They can significantly affect the speed, trajectory, and energy consumption. They play a pivotal role in underwater vehicles' dynamics and navigation. Incorporating ocean current models enables us to evaluate the intricate interplay between the vehicle's intended path and the prevailing oceanic flows.
We modified the mathematical model eq. \eqref{evect} to include the effect of ocean currents and given by,
\begin{equation}
     \boldsymbol{M\dot{\nu_r} +  C(\nu_r) \nu_r + D(\nu_r) \nu_r + g(\eta) = \tau}  \label{evectoc} 
\end{equation}
where, the relative velocity $\boldsymbol{\nu_r = \nu - \nu_c} = \left[\begin{array}{c} u - u_c~v-v_c~w-w_c~p~q~r \end{array}\right]^T = \left[\begin{array}{c} u_r~v_r~w_r~p~q~r \end{array}\right]^T$ and the ocean component $u_c,~v_c,~w_c$ is given by,
\begin{equation}
    \begin{split}
    & u_c = U_c\cos{\alpha_c}\cos{\beta_c} \\
    & v_c = U_c\sin{\beta_c} \\ \label{eoceancurrent}
    & w_c = U_c\sin{\alpha_c}\cos{\beta_c}    
    \end{split}
\end{equation}
Here, $U_c$ ocean current speed with its direction relative to the AUV is expressed in two angles: angle of attack $\alpha_c$ and side slip angle $\beta_c$. The model in eq. \eqref{eoceancurrent} is a 3-D ocean current model. The ocean current velocity is usually modelled as a first-order system with Gaussian white noise input $\eta$ \cite{Fossen2011}:
\begin{equation}
    \dot{U}_c + \zeta U_c = \mu \label{eocv}
\end{equation}
We choose $\zeta \geq 0$ such that the eq. \eqref{eocv} results in a stable system. Apart from $U_c$, we can associate dynamics to $\alpha_c$ and $\beta_c$. 

\section{Level Flight Dynamics and Linearization} \label{strim}

\subsection{Level flight dynamics}

Level flight is a condition where a constant set of inputs is given to the vehicle without any disturbances and the vehicle will maintain fixed velocity and orientation. It is also known as {\it trim} condition. It is a crucial process that involves adjusting the vehicle's configuration and control settings to achieve a state of equilibrium and stability in the water. An AUV when it is below the water surface has many forces acting on it simultaneously. In order for an AUV to satisfy the level flight conditions as mentioned earlier, all these forces and moments need to be balanced \cite{HA2011}, which will further result in zero acceleration and zero rates of rotation about all three axes. Mathematically this becomes a problem of estimating the control parameters which will result in this state. The control surfaces here are, stern angles, rudder angles and propeller 
   \[ \boldsymbol{u_i} = \left[ n ~~\delta_{s_1} ~~\delta_{s_2} ~~\delta_{r_1}~~ \delta_{r_2}\right]^T \]
   
The above vector consists of propeller RPM, two stern angles and two rudder angles. In order to solve for trim, the implementation of a mathematical model of the vehicle is done in the form of a function taking vehicle states along with control inputs and giving the rate of change of all the vehicle states as output. 
That means except horizontal position all other states will have their derivatives zero. 
\[\boldsymbol{\Omega} = [ \begin{array}{cccccccccccc}
   \dot{U} & \dot{\alpha}  & \dot{\beta} & \dot{p} & \dot{q} & \dot{r} & \dot{z} & \dot{\phi} & \dot{\theta} & \dot{\psi} & \dot{n} & \dot{{u_p}}
\end{array}]^T\]
 Here $U$ is the speed of the vehicle expressed as $\sqrt{u^2+v^2+w^2}$, $\alpha$ is the angle of attack, $\beta$ is the side-slip angle of the vehicle and ${u_p}$ is the propeller inflow velocity. Since the vehicle is not rotating, the body frame rotation rates are zero, i.e. $\dot{\phi} = \dot{\theta} = \dot{\psi} = p = q = r = 0 $. This implies the states in $\Omega$ should become zero. To achieve this optimum variables are chosen to be decision variables. In this case, the variables are, 
\[ \boldsymbol{d}= [ \begin{array}{cccccccccc}
\alpha & \beta & \theta & n & {u_p} & \tau & \delta_{s_1} & \delta_{s_2} & \delta_{r_1} & \delta_{r_2} 
\end{array}]^T\]

There are two ways of estimating these trim states. One is using Numerical methods like Newton Raphson. Another way is an analytical method wherein we put certain assumptions and based on that estimate expressions for all decision variables. After putting conditions the reduced 6 DOF equations will become 
\begin{eqnarray}
    (B-W) \theta+X_{u|u|} U^2+\left(1-\tau_p\right) T_{n|n|} n^2  &=&   0  \\ 
 (W-B) \phi+Y_{u v} U v+Y_{v|v|} v^2+Y_{u u \delta_r} U^2  (\delta_{r1}+\delta_{r2})   &=&0  \\
(W-B)\left[1-0.5 \theta^2-0.5 \phi^2\right]-Z_{w|w|} U^2 \theta^2+Z_{u w} U \theta+Z_{u u \delta_s} U^2 (\delta_{s1}+\delta_{s2})  &=& 0 \\
-z_g W \phi+Q_{n|n|} n^2 + {K_{roll}\delta_{roll}} &=& 0   \label{etrimfinal} \\
z_g\left(W+M_{u w} U^2\right) \theta-M_{w|w|} U^2 \theta^2+M_{u u \delta_s} U^2 (\delta_{s1} + \delta_{s2})&=& 0 \\
N_{u v} U v+N_{v|v|} v^2+N_{u u \delta_r} U^2 (\delta_{r1}+\delta_{r2})&=&0 
\end{eqnarray}
Furthermore, we address the trim scenario, wherein the objective is to nullify the roll motion. This can be effectively achieved through the utilization of decoupled control surfaces. By independently manipulating the two sterns ($\delta_{s_1}, \delta_{s_2}$) and two rudders ($\delta_{r_1}, \delta_{r_2}$), distinct adjustments can be made to ensure that the roll angle $\phi$ is minimized. Consequently, this configuration results in five essential control inputs to maintain a level flight condition: four control surface settings ($\delta_{s_1}$, $\delta_{s_2}$, $\delta_{r_1}$, and $\delta_{r_2}$), alongside a singular RPM value ($n_{prop}$).

\subsection{Linearization} \label{slin}

Linearization of an AUV involves approximating its nonlinear dynamic model with a simplified linear model around a specific operating point or reference condition. This linearization process is often used for control system design, stability analysis, and controller tuning. We consider linearization in two planes: one in the vertical plane known as the depth plane or dive plane, while the other is the horizontal plane known as the heading or steering subsystem. We would  These two subsystems involve approximating the nonlinear dynamic behaviour of the AUV's depth (vertical motion) and heading (yaw rotation) around a specific operating speed $U$.   

\subsubsection{Depth - Pitch Subsystem:}

In the depth plane, linearization focuses on understanding how changes in control surfaces $\delta_{s_1}$ and $\delta_{s_2}$, impact the AUV's vertical position. Since sterns are identical we have $\delta_s = \delta_{s_1} + \delta_{s_2}$ assuming no change in roll $\phi$. We consider the motion of the vehicle in the vertical plane with body-relative heave velocity w, the pitch rate q, the earth-relative vehicle depth z, and pitch angle ${\theta}$. Setting to zero the all unrelated terms $(v, p, r, \phi, y_{g}, z_g)$  in eq. \eqref{ekin}, eq. \eqref{e6dof1}, eq. \eqref{e6dof2}, eq. \eqref{force_equations} and eq. \eqref{moment_equations}, we get
\begin{eqnarray}
  \dot{z} &=&  -\sin{\theta}u + \cos{\theta}w  \\
    \dot{\theta}  &=&  q \\
  m(\dot{w} - uq -z_{g}q^{2}-x_{g}\dot{q})  &=& Z\\  \label{edpdyn1}
I_{yy}\dot{q} + m[z_{g}(\dot{u}+wq)-x_{g}(\dot{w}-uq)] &=& M
\end{eqnarray}
where, 
\begin{equation}
\begin{split}
  & Z = Z_{HS} + Z_{w|w|}w|w| + Z_{q|q|}q|q| + Z_{\dot{w}}\dot{w} + Z_{\dot{q}}\dot{q} + Z_{uq}uq  + Z_{uw}uw + Z_{{\delta_s}}\delta_s  \\
  &  M = M_{HS} + M_{w|w|}w|w| + M_{q|q|}q|q| + M_{\dot{w}}\dot{w} + M_{\dot{q}}\dot{q} + M_{uq}uq  + M_{uw}uw + M_{{\delta_s}}\delta_s
    \end{split}
\end{equation}
These equations can be linearized by assuming the vehicle perturbations around a steady point $U$. The speed $U$ in this case represents the steady-state forward velocity of the vehicle; the pitch is linearized about zero. As heave velocity is usually small we set $w=0$. Finally, substituting $u = U$ and linearized velocities and dropping higher-order terms \cite{Pres2001} results in the following linearized equations of motion,

\begin{equation}
\begin{split}
 &    \dot{z} = -\theta U + w  \\
 &   \dot{\theta} = q \\
&(I_{yy} - M_{\dot{q}})\dot{q} + (mx_gU - M_{q})q - M_{{\theta}}\theta   =  M_{{\delta_s}}\delta_s \end{split}
\end{equation}
where $M_{\dot{q}}$ is the added mass and the linearized coefficients are given as follows,
\begin{equation}
\begin{split}
    M_{\theta} & = -z_gW \\ \label{edplinear}
    M_{\delta_s} & = M_{uu\delta_{s}}U^2 \\
    M_q & = M_{uq}U
\end{split}
\end{equation} 
and $M_{uq} = M_{uqa} + M_{uqf}$ is the combined term with respect to added mass and fin lift coefficient.
The equations of motion can be further written down in matrix form in terms of a \textcolor{red}{three}-state vector containing \textcolor{red}{$(q, \theta, z)$}. Assuming that the heave velocities are small compared to the other terms, we can further reduce the equations of motion to this form:
\begin{equation}
\boldsymbol{\dot{x} = A_Dx + B_D}{\delta}_s \label{essdp}
\end{equation}
where,
\[\boldsymbol{x} = \left[\begin{array}{ccc}
     q  \\ \theta \\ z 
\end{array}\right]  ~~~ 
{\boldsymbol{A_D} = \left[\begin{array}{ccc}
\ \frac{M_q}{I_{yy}-M_{\dot{q}}} & -\frac{M_{\theta}}{I_{yy}-M_{\dot{q}}} & 0 \\ 
1 & 0 & 0 \\ 
0 & -U & 0
\end{array} \right]} ~~~ 
\boldsymbol{B_D} = \left[\begin{array}{c}
\frac{M_{\delta_{s}}}{I_{yy}-M_{\dot{q}}} \\ 0 \\ 0
\end{array}\right]\]

The transfer function of the vertical plane subsystem with input $\delta_{s}$ and output $z$ is,
\begin{equation}
    G_z(s)  \defi  \frac{z(s)}{\delta_s(s)}  =  -\frac{U}{s} \frac{\frac{M_{\delta_s}}{I_{yy}-M_{\dot{q}}}}{\left( s^2 - \frac{M_q}{I_{yy}-M_{\dot{q}}}s - \frac{M_{\theta}}{I_{yy}-M_{\dot{q}}} \right)} \label{etfdp}
\end{equation}
Further, the inner pitch loop transfer function relating the stern plane angle and pitch can be expressed as,
\begin{equation}
    G_{\theta}(s)  \defi  \frac{\theta(s)}{\delta_s(s)}  =  \frac{\frac{M_{\delta_s}}{I_{yy}-M_{\dot{q}}}}{\left( s^2 - \frac{M_q}{I_{yy}-M_{\dot{q}}}s - \frac{M_{\theta}}{I_{yy}-M_{\dot{q}}} \right)} \label{etfdp1}
\end{equation}
and the outer loop depth transfer function that relates vehicle pitch $\theta$ and the depth $z$ is,
\begin{equation}
    G_z(s)  \defi  \frac{z(s)}{\theta(s)}  =  -\frac{U}{s} \label{etfdp2}
\end{equation}


\subsubsection{Heading/Yaw Subsystem:}
In the heading subsystem, linearization involves approximating the effects of control surfaces on the AUV's yaw rotation. This is crucial for maintaining a desired heading or direction of travel. Linearized models in the heading plane allow for the design of control algorithms that enable the AUV to accurately follow desired paths, perform turns, or adjust its orientation in response to external factors. Automatic steering or heading control can be done using a rudder $\delta_{r_1}$ and $\delta_{r_2}$.  Since sterns are identical we have $\delta_r = \delta_{r_1} + \delta_{r_2}$.

The motion of the vehicle in the horizontal plane is considered. The body-relative sway velocity v, the yaw rate r, the earth-relative yaw angle ${\psi}$. All other velocities $(w, p, q)$ are set to zero linearize around the operating speed $U$ and substituting $u = U$. We set to zero all unrelated terms $(w, p, q, y_{g})$ in eq. \eqref{e6dof1}, eq. \eqref{e6dof2}, eq. \eqref{force_equations} and eq. \eqref{moment_equations}. We obtain the following equations,
\begin{equation}
\begin{split}
& \dot{\psi} = r \\
& m(\dot{v} + ur + x_{g}\dot{r}) = Y \\
 & I_{zz}\dot{r} + m[x_{g}(\dot{v} + ur)] = N
\end{split}
\end{equation}





Substituting $u = U$ and linearized velocities, and coefficients results in the following equations of motion,
\begin{equation}
\begin{split}
 &  \dot{\psi} = r \\
  &    (m-Y_{\dot{v}}\dot{v}) + (mx_{g} - Y_{\dot{r}})\dot{r} - {Y_{uv}U}v - ({Y_{ur}} - m)Ur = {Y_{uu\delta_r}}U^{2} \delta_r \\
&    (mx_{g} - N_{\dot{v}})\dot{v} + (I_{zz} - N_{\dot{r}})\dot{r} - {N_{uv}Uv - N_{ur}}Ur = {N_{uu\delta_r}}U^2 \delta_r
\end{split}
\end{equation}
Further, we assume the sway velocity is small $v = \dot{v} = 0$
\begin{equation}
\begin{split}
  &  \dot{\psi} = r \\
&   (I_{zz} - N_{\dot{r}})\dot{r} - (N_{r} - m x_{g}U)r = N_{{\delta_r}} \delta_r
\end{split}
\end{equation}
The corresponding two state $(r, \psi)$ state-space representation,
\begin{equation}
    \boldsymbol{\dot{x} = A_Hx+ B_H}{\delta}_R
\end{equation}
where,
\[\boldsymbol{x} = \left[\begin{array}{c}
    r \\ \psi
\end{array}\right]  ~~~
\boldsymbol{A_H} = \left[\begin{array}{ccc}
\frac{N_r}{I_{zz}-N_{\dot r}} & 0 \\ 
 1 & 0
\end{array}\right] ~~~
\boldsymbol{B_H} = \left[\begin{array}{c}
 \frac{N_{\delta_r}}{I_{zz}-N_{\dot r}}  \\ 0 
\end{array}\right]\]
The transfer function with $\delta_r$ as input and $\psi$ as output is given by,
\begin{equation}
G_{\psi}(s) \defi  \frac{\psi(s)}{\delta_r(s)} = \frac{\frac{N_{\delta_r}}{I_{zz}-N_{\dot{r}}}}{\left(s^2-\frac{N_{r}}{I_{zz}-N_{\dot{r}}}\right)}
\end{equation}
Further, we have the following relationship between $y$ and $\psi$,
\begin{equation}
   \dot y = U \psi
\end{equation}
Hence the transfer function,
\begin{equation}
  G_{y}(s) \defi \frac{y(s)}{\delta_r(s)} = \frac{U}{s} \frac{\frac{N_{\delta_r}}{I_{zz}-N_{\dot{r}}}}{s^2-\frac{N_{r}}{I_{zz}-N_{\dot{r}}}}  
\end{equation}
where $N_{\dot{r}}$ and $N_{\delta_r}$ are respectively the added mass and the fin lift coefficient.

\subsubsection{Speed Subsystem}

Only the surge equation of motion is considered to simplify the speed subsystem. We assume that the sway, heave, pitch, roll, and yaw coupling is insignificant.  Consequently, the surge equation can be simplified by merely retaining the terms referring to acceleration $\dot{u}$. We will set to zero all other velocities $(v, w, p, q, r)$, and drop the equations for any other direction.
\begin{equation}
(m - X_{\dot{u}})\dot{u} = X_{u|u|}u|u| + X_{prop}    
\end{equation}
where $X_{u|u|}$ is the axial drag and $X_{prop}$ is the thrust force due to the propeller in the X-direction.
Using trigonometric small-angle approximation, we arrive at the following result with $U$ as the surge velocity of the vehicle and substituting linearized axial drag $X_u$ and $X_{prop}$, we get:
\begin{equation}
(m - X_{\dot{u}})\dot{u} = X_u u + (1-\tau)T_{nn}n|n|    
\end{equation}
where $n$ is the angular velocity of the propeller. This equation is non-linear, so we need to linearize the $\Bar{n}^2$ term. This is accomplished by fitting a slope to the parabolic angular-velocity curve.
\[{n}^2 = m_{n}n\] with $m_{n}$ is a coefficient of the polynomial. 
The transfer function for the speed subsystem,
\begin{equation}
    G_u(s) \defi \frac{u(s)}{n(s)} = \frac{m_n(1-\tau)T_{nn}}{(m - X_{\dot{u}})s - X_u }\label{eq_lin_speed}
\end{equation}
We develop the control architecture in the preceding section for the horizontal plane (Heading), vertical plane (Depth - pitch) and speed subsystems.

\section{Simulation Results} \label{ssim}

Control systems are paramount in ensuring the AUV's autonomy and mission success. The design of control algorithms relies on the mathematical model of the AUV to determine appropriate control inputs that steer the vehicle to desired positions or trajectories. These control systems can be based on classical control techniques or modern approaches such as conventional controllers or SMC.

\subsection{Linear Controllers}

This section presents a comprehensive study of the linear controller architecture for the three subsystems: horizontal plane, vertical plane, and speed subsystem. The simulation results for these are also presented. 

\subsubsection{Depth-pitch controller:}
\begin{figure}[t]
    \centering
    \includegraphics[width = 15cm, height = 4cm]{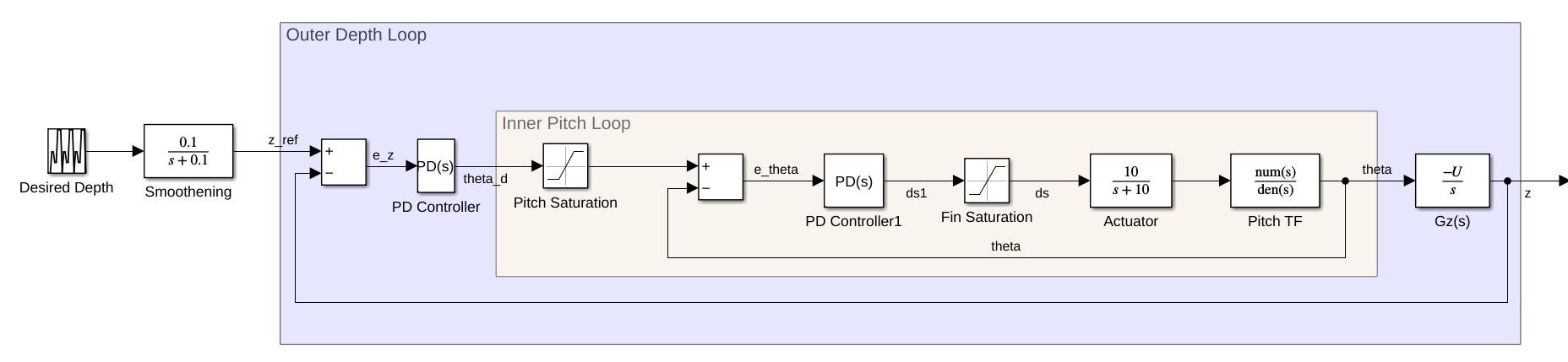} 
    \caption{Schematic of Depth - pitch subsystem with controller}
    \label{figdpss}
\end{figure}

The depth control system of a linearized AUV is a critical component responsible for maintaining the desired depth and pitch orientation during its underwater operations. The schematic of the depth control system consists of an outer depth loop and an inner pitch loop as shown in Figure \ref{figdpss}. The transfer function of the subsystem with $U$ as 4 knots is,
\begin{equation}
    G_{z}(s) = \frac{4.154}{s(s^2+0.824s+0.6927)}
\end{equation}


\begin{figure}
\centering
 \includegraphics[width = 10cm, height = 7cm]{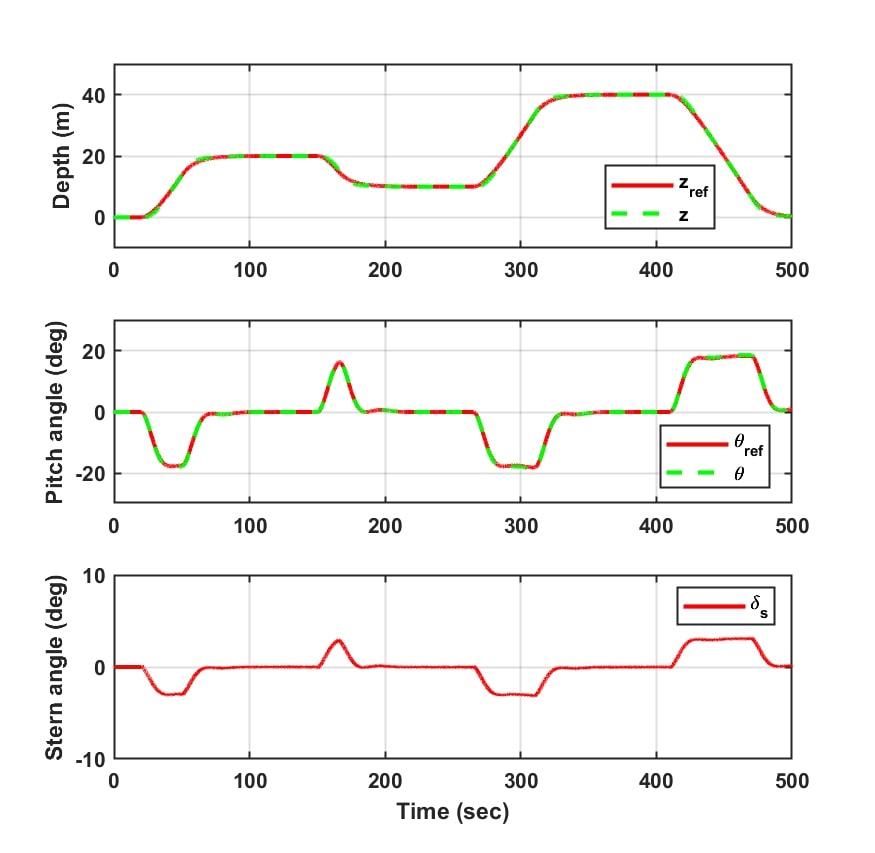}  
 \caption{Depth, pitch and stern angle of depth subsystem} \label{fdpres}
\end{figure}

For the outer depth loop, the required depth command is passed through a first-order low-pass filter to facilitate the refinement and smoothness of the curve. Eventually, the error between the desired depth and the actual depth is fed to a Proportional-Derivative (PD) controller which gives the desired pitch angle as output. The outer depth loop provides the desired pitch angle $\theta_{ref}$ and the control law is,
\begin{equation}
    \theta_{ref} =  -\left(K_{pd}e_{z} + K_{dd}\frac{d}{dt}e_{z}\right) \label{edcl} 
\end{equation}
where $e_{z} = z_{ref} - z$ error between the desired ($z_{ref}$) and actual depth ($z$), 
$K_{pd}$ and $K_{id}$ are the inner loop proportional and derivative gains respectively. Subsequently, the PD controller takes the error between this desired pitch angle and the actual pitch angle as input and outputs the required stern angle command. 
The control law for the inner pitch loop can be expressed as:
\begin{equation}
    \delta_s =  -\left(K_{pp}e_{\theta} + K_{dp}\frac{d}{dt}e_{\theta}\right) \label{epcl} 
\end{equation}
where $\delta_S$ is the required stern angle and $e_{\theta} = \theta_{ref} - \theta$ is the error between the required pitch angle $\theta_{ref}$ and actual pitch angle $\theta$, $K_{pp}$ and $K_{dp}$ are the outer loop proportional and derivative gains respectively.
The simulation results of the depth-pitch subsystem with the controller are shown in Figure \ref{fdpres}.The reference signal for the desired depth (\(z_{ref}\)) is depicted as a solid red line and serves as the input reference for the system. The first subplot presents a comparative analysis of the desired and actual depth profiles over a specified time span. This visual comparison provides valuable insights into the depth control performance of the controller, illustrating how effectively the controller responds to dynamic changes in depth references. Moving to the second subplot, a clear representation of the vehicle's pitch angle response within the influence of the applied control system is observed. The plots of the \(\theta_{ref}\) and \(\theta\) over time indicate that the controller accurately tracks the desired pitch angle, showcasing its precision. The third subplot, depicted in Figure \ref{fdpres}, illustrates the evolution of the required stern angle command over time in response to varying desired depth commands. Notably, due to the small magnitude of the depth change, the maximum required stern angle remains around 5 degrees. This analysis effectively demonstrates the controller's ability to accurately follow the desired depth trajectory.


\subsubsection{Yaw Controller:}
\begin{figure}
    \centering
    \includegraphics[width = 15cm, height = 3.5cm]{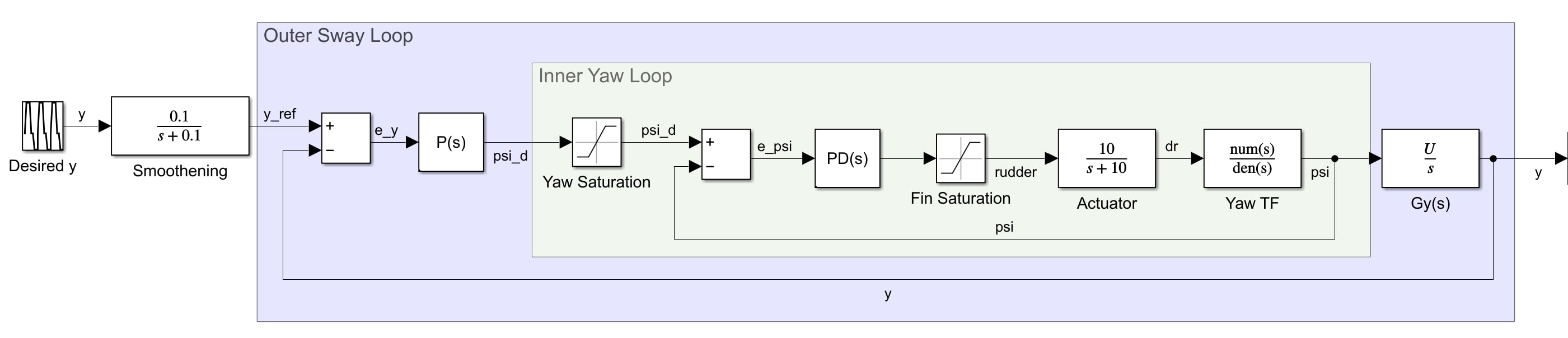}
    \caption{Schematic of yaw subsystem with controller}
    \label{figyss}
\end{figure}

We follow a similar architecture for yaw control of the vehicle. The schematic of the yaw control is shown in Figure \ref{figyss}. There are two loops as before, the inner yaw loop and the outer y direction control loop. The transfer function of the yaw subsystem is,
\begin{equation}
    G_{\psi}(s) = -\frac{4.223}{s(s+0.579)}
\end{equation}
The control law for the inner loop is given by,
\begin{equation}
    \delta_r =  -\left(K_{p\psi}e_{\psi} + K_{d\psi}\frac{d}{dt}e_{\psi}\right) \label{eycl} 
\end{equation}
where $\delta_r$ is the required rudder angle to provide the required yawing of the vehicle and $e_{\psi} = \psi_{ref} - \psi$. $K_{p\psi}$ and $K_{d\psi}$ are the inner loop controller gains. The outer loop provides the desired yaw angle $\psi_{ref}$ and the control law is,
\begin{equation}
    \psi_{ref} =  -\left(K_{py}e_{y} + K_{dy}\frac{d}{dt}e_{y}\right) \label{eyacl} 
\end{equation}
where $e_{y} = y_{ref} - y$ error between the desired and actual $y$ and $K_{py}$ and $K_{dy}$ are the outer loop controller gains.


\begin{figure}
    \centering
    \includegraphics[width = 10cm, height = 6cm]{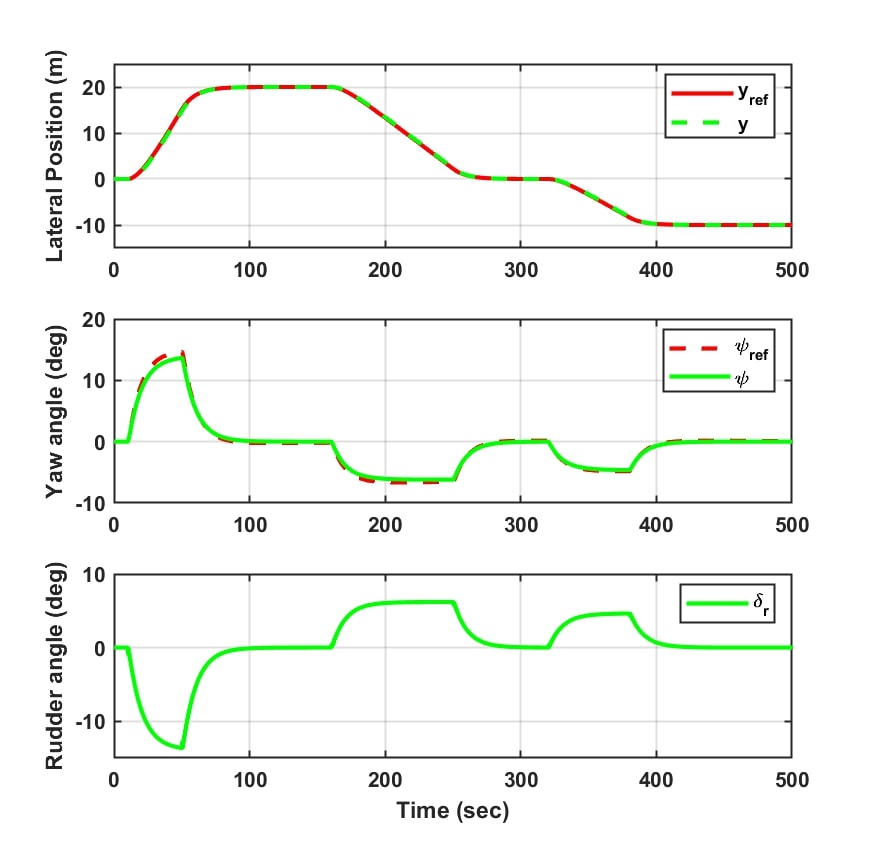}
    \caption{Lateral position, yaw and rudder angle of yaw subsystem}
    \label{lin_yaw_res}
\end{figure}

In a similar fashion as the depth subsystem, a signal of varying $y$ is sent to the outer controller as a reference and the desired rudder angle is obtained for the same. The results of a similar manoeuver are given in Figure \ref{lin_yaw_res}. Here the first subplot represents the desired lateral position($y_{ref}$) and actual lateral position($y$). It is seen that the desired position is being followed precisely. The respective yaw angles are given in the second subplot. The maximum value of the yaw angle depends on the value of $y_{ref}$ as well as the time taken to reach the same. As seen from the figure, in this manoeuvre for the demand of zero to 20 meters in around one minute at a constant speed of 3 knots a maximum of 15 degrees of yaw angle is required which is being followed accurately. The third subplot gives the desired rudder angle to obtain the yaw requirement for the given manoeuvre. 


\subsubsection{Speed subsystem:}
\begin{figure}[t]
    \centering
    \includegraphics[width = 12cm, height = 2.5cm]{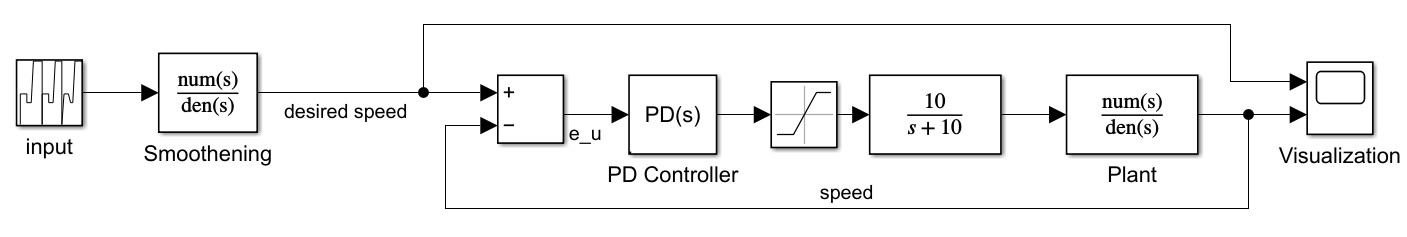}  
    \caption{Schematic of speed subsystem}
    \label{lin_speed_arch}
\end{figure}

The presented simulation results offer a comprehensive portrayal of the speed controller's tracking capabilities. Through a comparative analysis, the required speed is contrasted with the actual speed profile of the vehicle, spanning a specified time interval. This comprehensive control system architecture is shown in Figure \ref{lin_speed_arch}, offering a tangible overview of the interactions between the components involved in regulating the AUV's speed.The desired speed command (\(u_{ref}\)) is compared with the actual speed (\(u\)). The difference between these two values is fed to the Proportional Derivative (PD) controller, which subsequently generates the necessary RPM output to achieve the specified speed. This RPM output then acts as an input to the system, as encapsulated by the transfer function outlined in eq. \eqref{eq_lin_speed}. The transfer function is,
\begin{equation}
    G_{u}(s) = \frac{0.038}{(31.41s+1.6)}
\end{equation}


The speed control system is designed by applying the following control law:
\begin{equation}
   n_{prop}(t) = K_{ps}e_{spe} + K_{ds}\frac{d}{dt}e_{spe} \label{espcl} 
\end{equation}
where $e_{u} = u_{ref} - u$, the error between the desired speed and the actual speed of the vehicle, $K_{ps}$ and $K_{ds}$ are proportional and derivative gains for speed controller respectively.

\begin{figure}
    \centering
    \includegraphics[width = 10cm, height = 8cm]{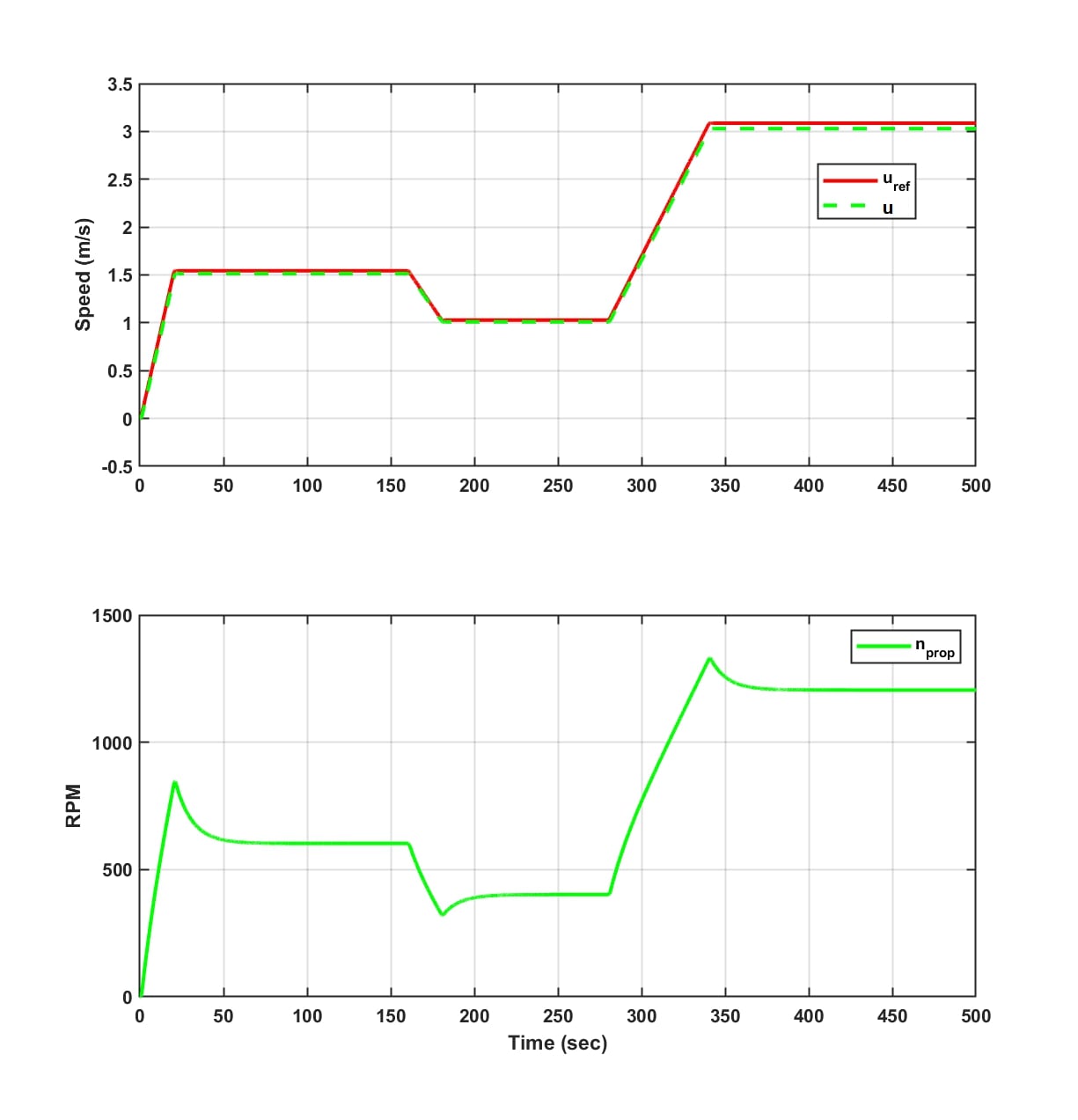}
    \caption{Speed and RPM of the vehicle}
    \label{fig_lin_speed_res}
\end{figure}

In addition to speed tracking response of the vehicle's propulsion system is examined in terms of motor RPM. By examining the change in required RPM values to maintain the desired speed we can gauge the control effort generated by the controller and understand the performance of the controller. Figure \ref{fig_lin_speed_res} depicts the outcomes of our linear simulations. The first subplot displays the desired and actual speed profiles, demonstrating the controller's performance in maintaining the desired speed. Here $u_{ref}$ is the desired speed and $u$ is the actual speed attained by the vehicle. The second subplot presents corresponding required RPM values $n_{prop}$, providing insight into the control effort exerted on the propulsion system. These results collectively showcase the effectiveness and robustness of the speed controller for varying demands.

\subsection{Level Flight Results}

This section presents the outcomes of AUV's level flight analysis. To achieve this, the nonlinear AUV model undergoes linearization at the operational velocity of $U = 4$ knots. The simulation incorporates specific prerequisites: a consistent speed profile ($\dot{u} = \dot{v} = \dot{w} = 0$), constant depth and heading ($z = \psi = \text{constant}$), as well as negligible rotation rates ($\dot{\phi} = \dot{\theta} = \dot{\psi} = 0$) and angular velocities ($p = q = r = 0$). The results of the level flight estimation for the research vehicle at 4 knots are summarized in Table \ref{ttrim}.
\begin{table} 
    \centering
    \renewcommand{\arraystretch}{1.5}
\setlength{\tabcolsep}{7pt} 
     \caption{Level flight of AUV at 4 Knots}
    \begin{tabular}{|c|c|c|c|c|}
    \hline Variable & Symbol & Units & Value \\
\hline Body-frame surge velocity & $u$ & $m s^{-1}$ & 2.0577   \\
\hline Body-frame sway velocity & $v$ & $m s^{-1}$ & 0.001   \\
\hline Body-frame heave velocity & $w$ & $m s^{-1}$ & -0.011   \\
\hline Earth-frame roll & $\phi$ & $deg$ & -2.5941 \\
\hline Earth-frame pitch & $\theta$ & $deg$ & -0.72 \\
\hline Propeller rotation rate & $n$ & $rpm$ & 1413  \\
\hline Angle of attack & $\alpha$ & $deg$ & -0.72  \\
\hline Angle of sideslip & $\beta$ & $deg$  & -0.0276  \\
\hline Propeller inflow rate & $u_p$ & $ ms^{-1}$  & 1.2936 \\
\hline Stern angle 1/ 2  & $\delta_{s_1}$/$\delta_{s_2}$ & $deg$ & -1.4  \\
\hline Rudder angle 1/2 & $\delta_{r_1}$ /$\delta_{r_2}$& deg & -0.05  \\
\hline
    \end{tabular}
       \label{ttrim}
\end{table}

\begin{figure}
    \centering
    \begin{tabular}{cc}
       \includegraphics[width = 7.5cm, height = 6cm]{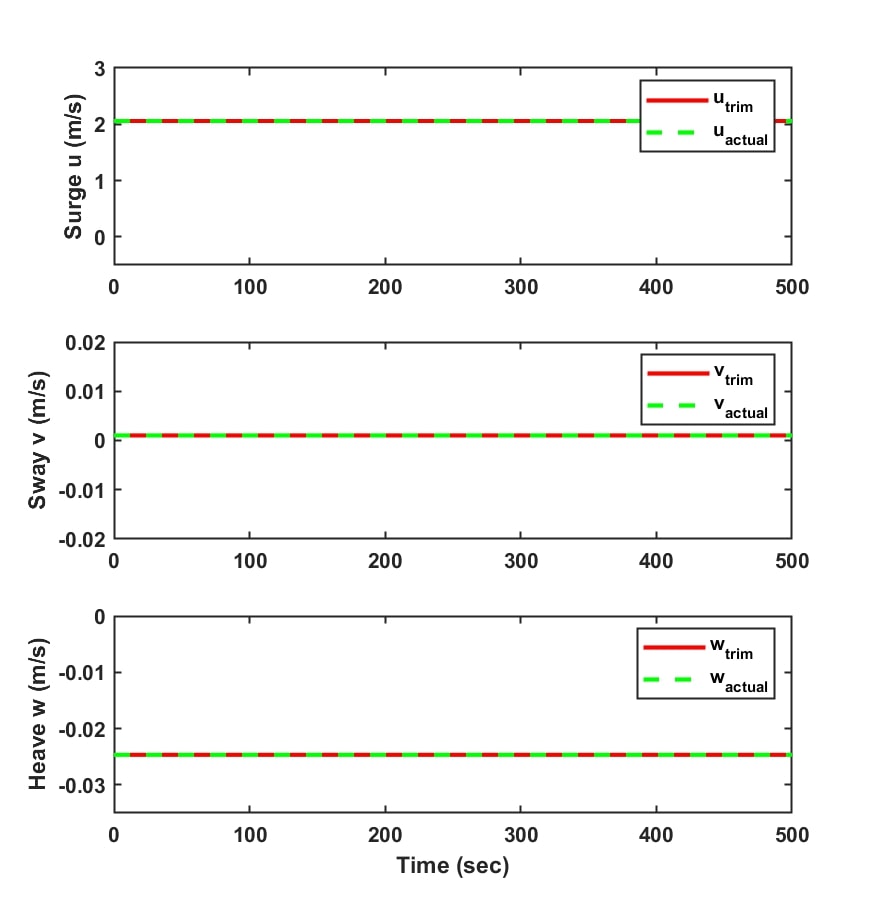} 
      &
        \includegraphics[width = 7.5cm, height = 6cm]{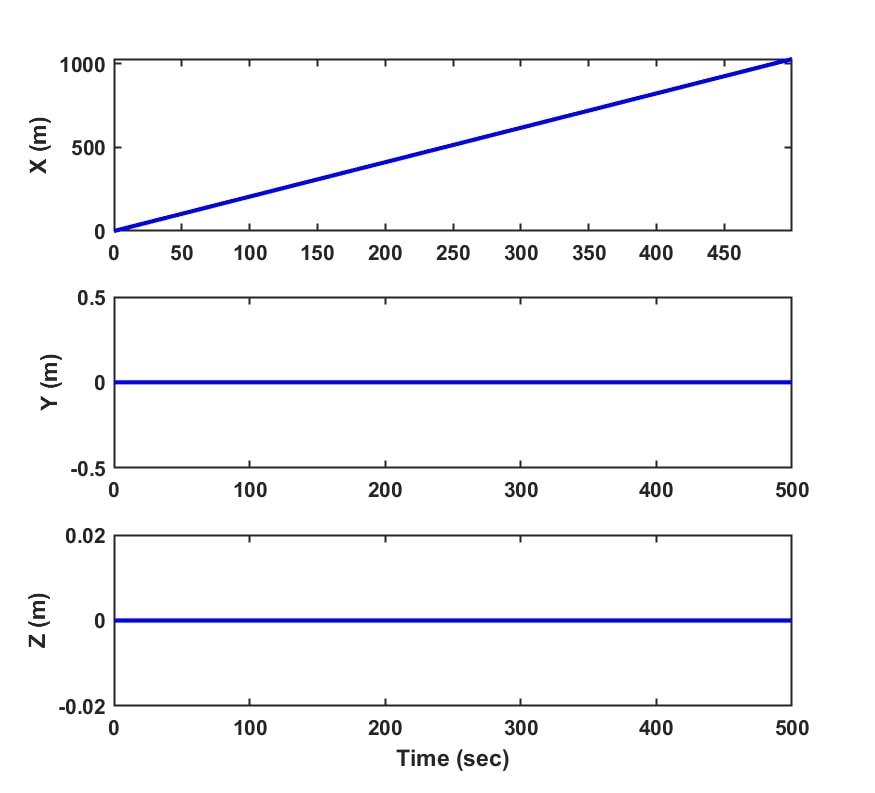} \\
        (a) & (b)
    \end{tabular}
    \caption{(a)~Body-frame linear velocities $u,~v$ and $w$. (b)~AUV motion in the earth-fixed frame.}
    \label{ftrim1}
\end{figure}

These level flight estimated values were given as initial condition along with the constant control input vector,
\begin{equation}
    \left[ \begin{array}{ccccc} n~\delta_{s_1}~\delta_{s_2}~\delta_{r_1}~\delta_{r_2} \end{array}\right]^T = \left[ \begin{array}{ccccc} 1413~-1.2648~-1.2648~-0.0561~ -0.0561 \end{array}\right]^T 
\end{equation}
\begin{figure}
    \centering
    \begin{tabular}{cc}
      \includegraphics[width = 7.5cm, height = 6cm]{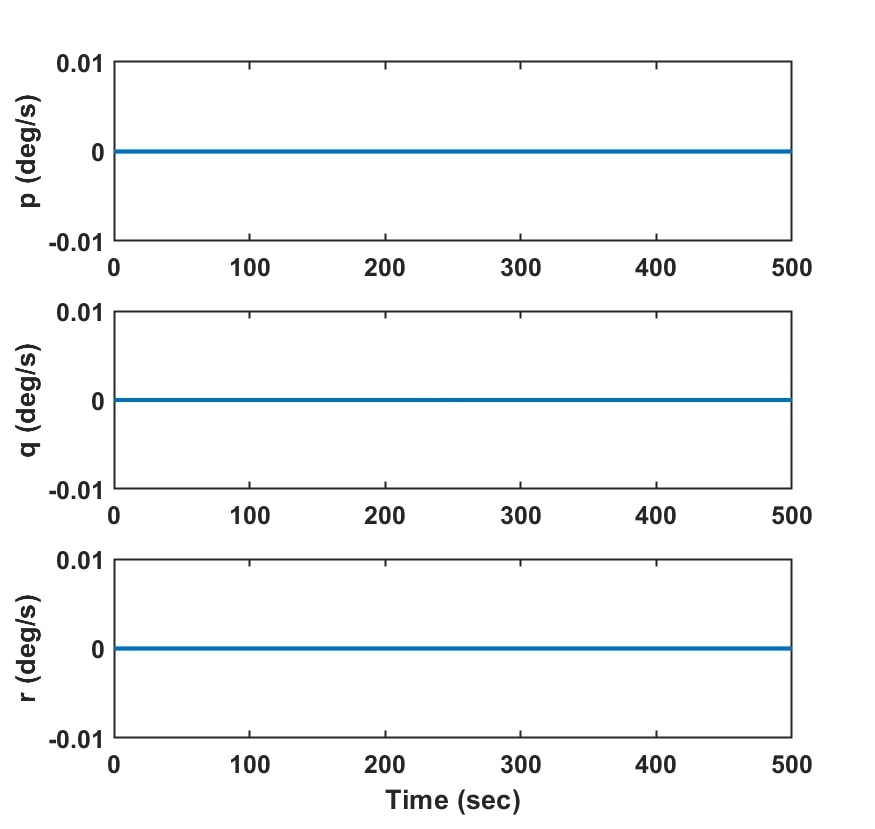}
       \includegraphics[width = 7.5cm, height = 6cm]{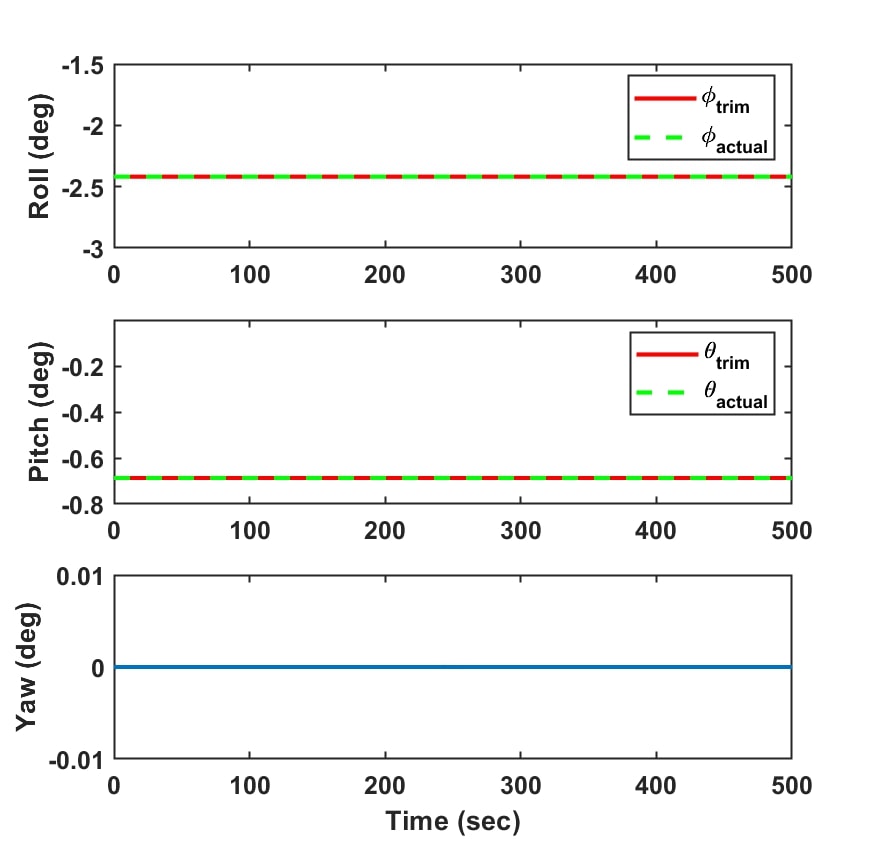} \\
       (a) & (b)
    \end{tabular}
    \caption{(a)~Body-frame angular velocities $u,~v$ and $w$. (b)~Earth-frame angular positions $\phi,~\theta$ and $\psi$.}
    \label{ftrim2}
\end{figure}

The control inputs $n$ are expressed in RPM, while the stern and rudder angles are in degrees. The trim values (shown as a solid red line) are compared against the actuals wherever relevant in Figures \ref{ftrim1} and \ref{ftrim2}. Figure \ref{ftrim1} (a) demonstrates that the AUV effectively sustains consistent velocities $u$, $v$, and $w$ as given in Table \ref{ttrim}. Additionally, the plot in Figure \ref{ftrim1} (b) depicts the persistent motion along the $x$-direction, while exhibiting no discernible variation along the $y$ and $z$ axes. Furthermore, Figure \ref{ftrim2} portrays the angular velocities, all of which are maintained at zero. The AUV's roll angle is stabilized at $-2.594^\circ$, while the pitch remains at $-0.72^\circ$. Notably, the yaw angle remains constant at $0^\circ$. It is evident that the vehicle maintains its steady state for the trim values given in Table \ref{ttrim}.

\subsection{Nonlinear Controller}

\begin{figure}[t]
    \centering
      \includegraphics[width = 13cm, height = 6cm]{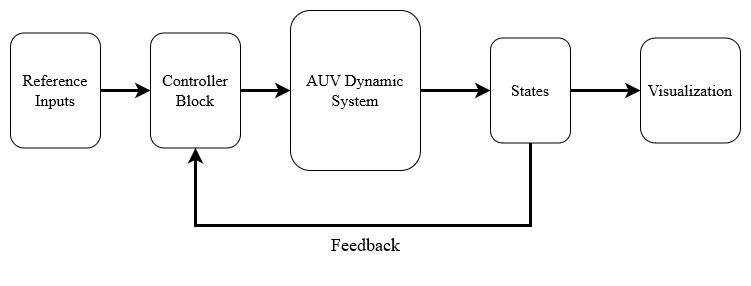} 
    \caption{Block diagram of Nonlinear AUV model}
     \label{figsche}
\end{figure}

The conventional controllers and SMC are implemented to achieve the desired speed, depth and steering. The nonlinear model is given in eq. \eqref{ematrix} is implemented in MATLAB-Simulink. The schematic of the research vehicle in Simulink is shown in Figure \ref{figsche}. The References subsystem serves as the source of desired values, encompassing speed, y-position, z-position, and roll angle. These desired values are subsequently transmitted as inputs to the controller block. Within the controller block, there exist four distinct controllers, each aligned with a specific reference signal. Each controller operates as an independent subsystem, employing the principles elucidated in the linear control section. Eventually, it calculates the requisite control commands needed for tracking. The controller block is either conventional controllers or SMC. Further, the AUV dynamics block takes the control inputs furnished by the controller block along with the state feedback information. These combined elements facilitate the estimation of the subsequent state of the AUV. 


This simulation model will be used for all our simulations. To demonstrate the efficacy of both conventional control methods and the Sliding Mode Controller (SMC), a series of simulations are conducted. The simulations are structured as follows:
Initially, the performance of a conventional controller is assessed. This evaluation encompasses scenarios involving variations in the center of gravity (CG) and the presence of ocean currents. The Conventional controller's response is observed under these conditions to gauge its effectiveness in maintaining stability and control. Subsequently, the focus shifts to the Sliding Mode Controller. The efficacy of SMC is demonstrated using lawn mowing in the presence of CG variations, ocean currents, and parameter variation ($X_{u|u}$ value changed by 50\%). By subjecting the SMC to these challenges, its robustness and capability to navigate and counteract disturbances are evaluated.

\subsubsection{Conventional Controller:}

A comparative assessment of the Conventional controller's performance is conducted under two distinct scenarios: with and without variation in the center of gravity (\(CG\)) location. This evaluation aims to discern the impact of \(CG\) location changes on the controller's effectiveness in regulating the system. In the first scenario, the Conventional controller is examined in a setting where the \(CG\) location remains constant. The system's response to input commands and disturbances is analyzed, providing insight into the controller's inherent capabilities in maintaining stability and precision.  In the second scenario, the Conventional controller's performance is investigated when confronted with variations in the \(CG\) location ($x_{cg} = 8mm$). This dynamic factor introduces an additional layer of complexity, potentially affecting the system's stability and control.

\begin{figure}[t]
    \centering
       \includegraphics[width = 12cm, height =10cm]{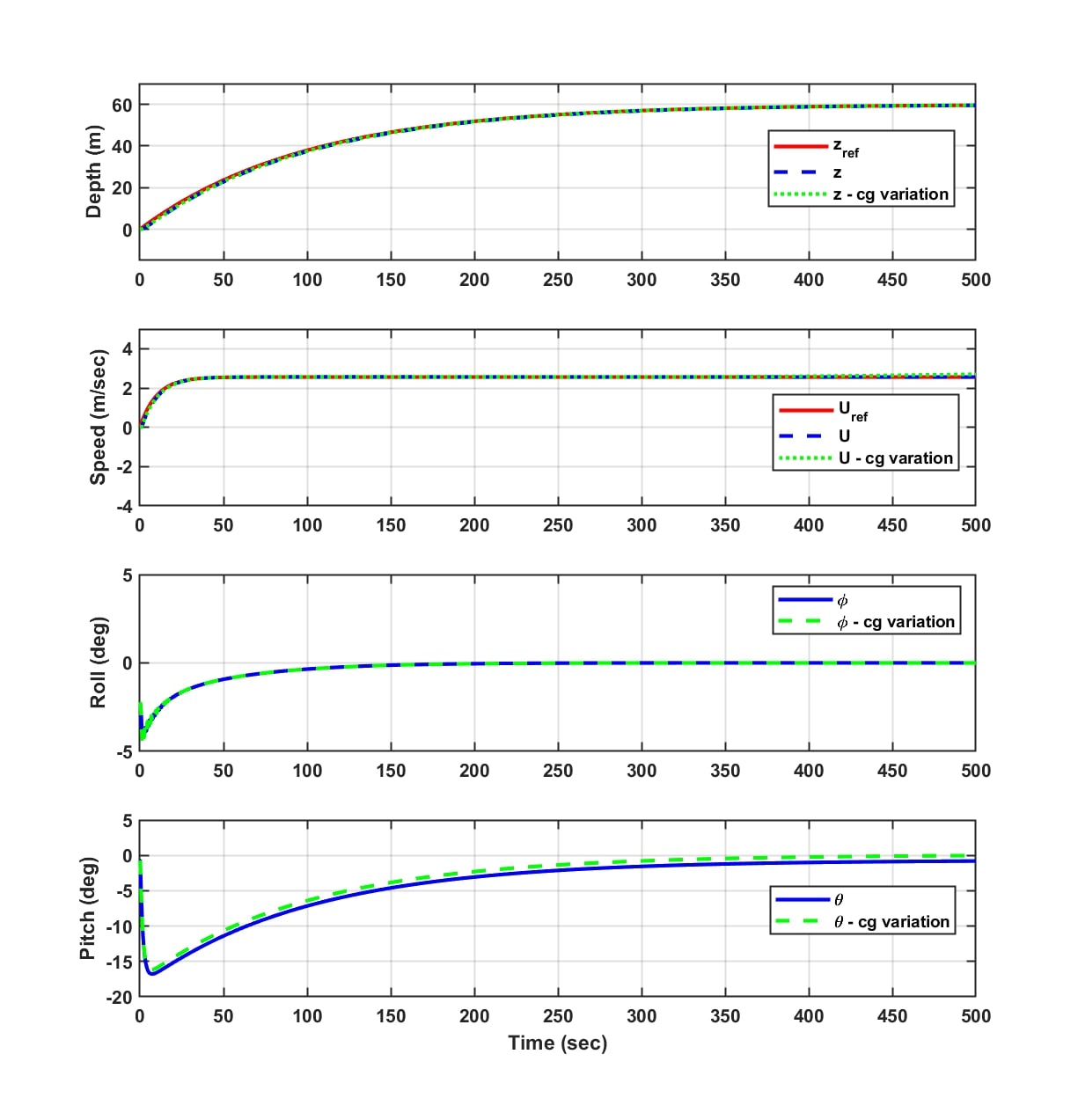} 
    \caption{Comparison of Conventional controller performance with and without variation in $CG$ location}
    \label{fig_xcg_var}
\end{figure}

Figure \ref{fig_xcg_var} shows a set of simulation data with and without the variation of the location of CG. For this simulation the references used are: $u_{ref}$ = 4 knots, $z_{ref}$ = 60m, $y_{ref}$ = 0 and $\phi_{ref}$ = 0. Here first subplot shows the variation of depths in both cases along with reference. The second subplot shows the variation of speed in both cases whereas the third and fourth show the roll and pitch angles respectively for both cases.

In order to effectively nullify the roll angle $\phi$, a strategic approach is employed as outlined below:

\begin{equation}
    \begin{split}
       & \delta_{s_1} = \delta_s + \kappa \frac{\delta_{a}}{2} \\
       & \delta_{s_2} = \delta_s - \kappa \frac{\delta_{a}}{2} \\ \label{erollcontrol}
       & \delta_{r_1} = \delta_r + (1- \kappa) \frac{\delta_{a}}{2} \\
       & \delta_{r_2} = \delta_r - (1-\kappa) \frac{\delta_{a}}{2} \\       
    \end{split}
\end{equation}

where, the parameter $\kappa$ assumes a pivotal role, determining the degree of emphasis attributed to the stern and rudder angles within the endeavor to nullify the roll angle and $\delta_a$ represents the requisite output emerging from the roll controller housed within the controller block, as depicted in Figure \ref{figsche}. This orchestrated strategy meticulously adjusts control inputs through the defined equations, aiming to achieve the specific objective of nullifying the roll angle. This is evidently shown in Figure \ref{fig_xcg_var}. The vehicle's roll angle $\phi$ gradually approaches zero from its initial value of $-2.59^\circ$. While a minor undershoot is observed during the initial phase, the roll angle eventually stabilizes at zero, demonstrating a successful convergence to the desired state.
Analyzing these outcomes collectively reveals that even with minor adjustments in $x_{cg}$, the system adeptly achieves the desired outcomes. It is noteworthy that slight alterations in trim values for specific variables might occur.
By contrasting the results with the first scenario, the Conventional controller's robustness and adaptability in handling \(CG\) location changes can be elucidated.

\begin{figure}
    \centering
      \includegraphics[width = 8cm, height = 8cm]{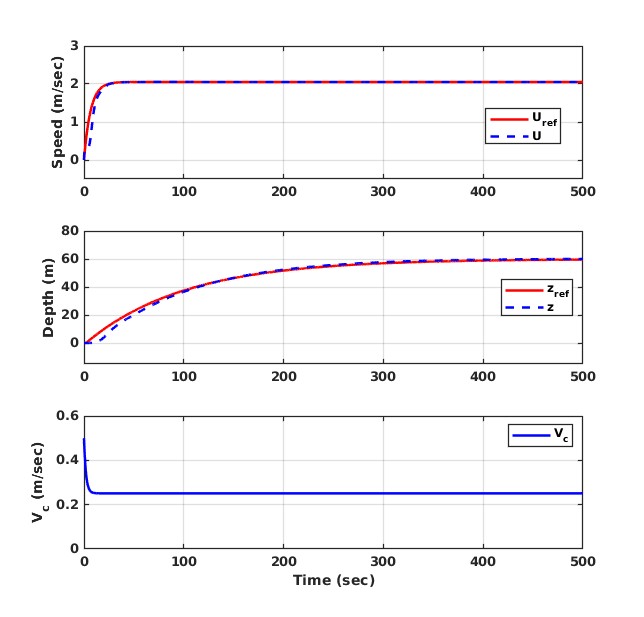} 
    \caption{Simulation results with ocean currents}
    \label{fig_oc}
\end{figure}

Further, we implement the ocean current model into the simulation framework. 
The ocean current model considered is simulated using eq. \eqref{eoceancurrent}. The ocean current relative velocity, $V_c$, is generated with Gaussian input, $\zeta = 1$ and initial condition $V_{c_0} = 0.5$. We have set $\alpha_c$ to $0.035 rad$ $(~2^\circ)$ and $\beta_c$ to $1.221 rad (~70^\circ)$.
This ocean current model mostly affects the speed and depth of the AUV. This model of ocean currents was added to the same simulations as performed earlier and results are shown in Figure \ref{fig_oc}.


The first subplot reveals a minute difference between the desired speed $U_{ref}$ and the actual speed of the vehicle $U$. Although there is a slight transient and persists for very few seconds. Later the controller seems to be successful in maintaining the desired speed. Similarly, the depth $z$ in the second subplot follows the desired depth command $z_{ref}$. The third subplot shows the variation of $U_c$ over time. It can be seen that $U_c$ begins with the initial value $0.5 m/s$ and reaches a steady state value of $0.25 m/s$. The simulations have showcased the effectiveness of conventional controllers in maintaining desired vehicle trajectories and depths even in the presence of ocean currents. More complex ocean models can be used to increase the accuracy of the simulation.

\subsubsection{Comparison of SMC and Conventional controller: Lawn Mowing Manoeuvre} 

In this section, we build a control system architecture where sliding mode controllers are used for the inner pitch and yaw control loops. The core idea behind sliding mode control is to shape a certain trajectory of system states as they advance over a defined manifold known as the sliding surface. Usually, the sliding surface depends on the tracking error (e). For inner loop pitch control $\theta$ is the state variable and for inner loop yaw control $\psi$ is the state variable. The sliding surfaces for pitch and yaw inner loops respectively are,
\begin{eqnarray}
    s_{pitch} & = \gamma_{\theta_1} \dot{e}_{\theta} + \gamma_{\theta_2}e_{\theta} & \label{eslidsurp} \\
    s_{yaw} & = \gamma_{\psi_1} \dot{e}_{\psi} + \gamma_{\psi_2} e_{\psi} & \label{eslidsury}    
\end{eqnarray}
where $e_{\theta}$ is the error between $\theta_{ref}$ and $\theta$ and $e_{\psi}$ error between $\psi_{ref}$ and $\psi$. The sliding mode controller design process involves tuning these constants $\gamma_{\theta_1}$, $\gamma_{\theta_2}$, $\gamma_{\psi_1}$ and $\gamma_{\psi_2}$ such that
the derivatives of respective control surfaces go to zero. 
\begin{eqnarray}
    \dot{s}_{pitch} & = 0 & \label{control_law_estimation_pitch} \\
    \dot{s}_{yaw}   & = 0 & \label{control_law_estimation_yaw}
\end{eqnarray}
By actively steering the system states toward and subsequently maintaining them on the sliding surface, sliding mode control offers a robust and dependable strategy well-suited to a spectrum of intricate and dynamically evolving systems. The control law for the inner pitch  loop
\begin{equation}
    u_{pitch} = \delta_{s} = \gamma_{\theta_3}sgn(s_{pitch})\label{eslidup}
\end{equation}
and for the yaw is given by,
\begin{equation}
      u_{yaw} = \delta_{r} = \gamma_{\psi_3}sgn(s_{psi})\label{esliduy}
\end{equation}
The control law in SMC is designed to drive the system's states to the sliding surface and maintain them there. However, a critical challenge in SMC is the occurrence of chattering, which refers to rapid and frequent switching between control actions near the sliding surface. Chattering can lead to undesirable high-frequency oscillations in the control signal, potentially affecting system performance and wear on actuators. To mitigate this, we use the saturation function in the control law. These functions can limit the rate of change of control inputs, preventing rapid fluctuations and reducing chattering.

\begin{figure}
    \centering
     \includegraphics[width = 12cm, height = 8cm]{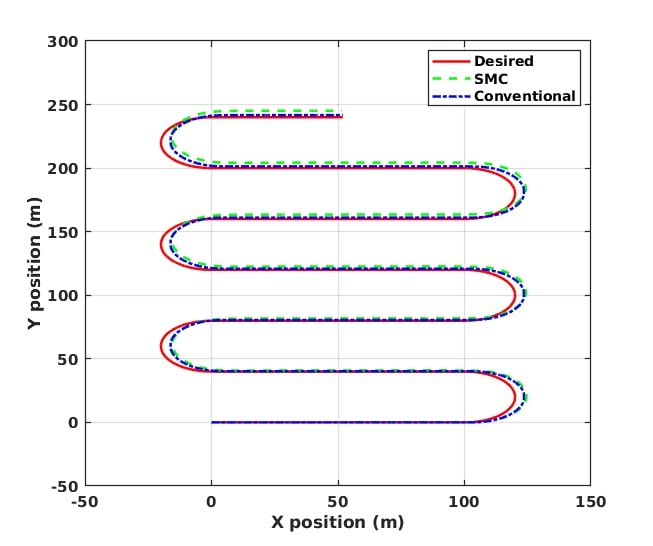} 
    \caption{Lawn mowing in XY plane with SMC and Conventional controller}
    \label{fig_smcpidlm}
\end{figure}

A classical lawn mowing manoeuvre is performed to showcase the ability of conventional and SMC controllers. This manoeuvre is essentially used for underwater object detection, search and rescue, environmental surveys and many other applications. It gives a broad sense of the ability of the vehicle to operate autonomously. The conventional controller architecture remains unchanged, serving as a benchmark for evaluation. In contrast, SMC system is introduced to handle the inner pitch and yaw control functions, following the equations eq. \eqref{eslidsurp}, eq. \eqref{eslidsury}, eq. \eqref{eslidup}, and eq. \eqref{esliduy}. A visual representation of the AUV's positional dynamics (in XY plane) is vividly presented through the plot in Figure \ref{fig_smcpidlm}. Evidently, both controllers successfully executed the planned manoeuvre. 


Further, the comparative analysis between the two controllers is conducted under the umbrella of uncertainties. In this experimental framework, the target speed is set at $U_{ref} = 4~Knots$ and the vehicle is assumed to be at the desired depth of $60 m$. To comprehend the system's behaviour in response to uncertainties, we subject the system to a series of desired commands that yield a lawn-mowing manoeuvre. Notably, this examination involves varying a single parameter at a time, allowing us to scrutinize the performance of both controllers. During this investigation, our focus was the error along the y-axis as there was minimal or no change in the x-direction, hence the root-mean-square error (RMSE) in the y-direction is recorded for each variation as
\begin{equation}
RMSE = \sqrt{\frac{(y_{ref} - y)^2}{N}},
\end{equation}
where, $y_{ref}$ is the desired value in the $y$-direction, $y$ is the actual value and $N$ is the number of the samples. 
This choice serves as a metric to assess the effectiveness of the controllers under the influence of parameter variations. To ensure clarity and precision, only one parameter is altered and results are recorded. The comparison of two controllers in terms of  (RMSE) are tabulated in Table \ref{coeff_var_lm}.
\begin{table}[]
\begin{center}
\renewcommand{\arraystretch}{1.5}
\setlength{\tabcolsep}{7pt} 
\caption{RMSE of Conventional and SMC due to variation in hydrodynamic coefficients}
\begin{tabular}{|c|ccccccccc|}
\hline
\multirow{2}{*}{Controller} & \multicolumn{9}{c|}{Coeffecient}             \\\cline{2-10} 
                            & \multicolumn{3}{c|}{$X_{u|u|}$}& \multicolumn{3}{c|}{$M_{uw}$}& \multicolumn{3}{c|}{$Z_{uw}$}\\ \hline
\% change                   & \multicolumn{1}{c|}{$-25\%$}  & \multicolumn{1}{c|}{$0\%$}   & \multicolumn{1}{c|}{$+25\%$} & \multicolumn{1}{c|}{$-25\%$}  & \multicolumn{1}{c|}{$0\%$}   & \multicolumn{1}{c|}{$+25\%$} & \multicolumn{1}{c|}{$-25\%$}  & \multicolumn{1}{c|}{$0\%$}   & $+25\%$ \\ \hline
Conventional                         & \multicolumn{1}{c|}{$0.506$} & \multicolumn{1}{c|}{$0.504$} & \multicolumn{1}{c|}{$0.504$} & \multicolumn{1}{c|}{$0.502$} & \multicolumn{1}{c|}{$0.504$} & \multicolumn{1}{c|}{$0.507$} & \multicolumn{1}{c|}{$0.503$} & \multicolumn{1}{c|}{$0.504$} & $0.507$ \\ \hline
SMC                         & \multicolumn{1}{c|}{$1.600$} & \multicolumn{1}{c|}{$1.614$} & \multicolumn{1}{c|}{$1.623$} & \multicolumn{1}{c|}{$1.613$} & \multicolumn{1}{c|}{$1.614$} & \multicolumn{1}{c|}{$1.611$} & \multicolumn{1}{c|}{$1.600$} & \multicolumn{1}{c|}{$1.614$} & $1.611$ \\ \hline
\end{tabular}\label{coeff_var_lm}
\end{center}
\end{table}
Similarly, the simulations are repeated for different values of buoyancy. The results are given in \ref{Buoyancy_var_lm}. As observed from Table \ref{coeff_var_lm} and \ref{Buoyancy_var_lm}, the error in the case of conventional is found to be lesser compared to SMC controller. When observed closely, in the case of SMC it can be seen that with parameter variation the change in error is minimal compared to conventional controller.
\begin{table}[]
\begin{center}
\renewcommand{\arraystretch}{1.5}
\setlength{\tabcolsep}{10pt} 
\caption{RMSE of Conventional and SMC due to variation in buoyancy}
\begin{tabular}{|c|cccc|}
\hline
\multirow{2}{*}{Controller} & \multicolumn{4}{c|}{(B-W)/g  (grams)}                                                        \\ \cline{2-5} 
                            & \multicolumn{1}{c|}{100}   & \multicolumn{1}{c|}{340}   & \multicolumn{1}{c|}{650}   & 1000  \\ \hline
Conventional                & \multicolumn{1}{c|}{0.513} & \multicolumn{1}{c|}{0.505} & \multicolumn{1}{c|}{0.495} & 0.483 \\ \hline
SMC                         & \multicolumn{1}{c|}{1.613} & \multicolumn{1}{c|}{1.614} & \multicolumn{1}{c|}{1.612} & 1.613 \\ \hline
\end{tabular}\label{Buoyancy_var_lm}
\end{center}
\end{table}

\begin{figure}
    \centering
      \includegraphics[width = 12cm, height = 8cm]{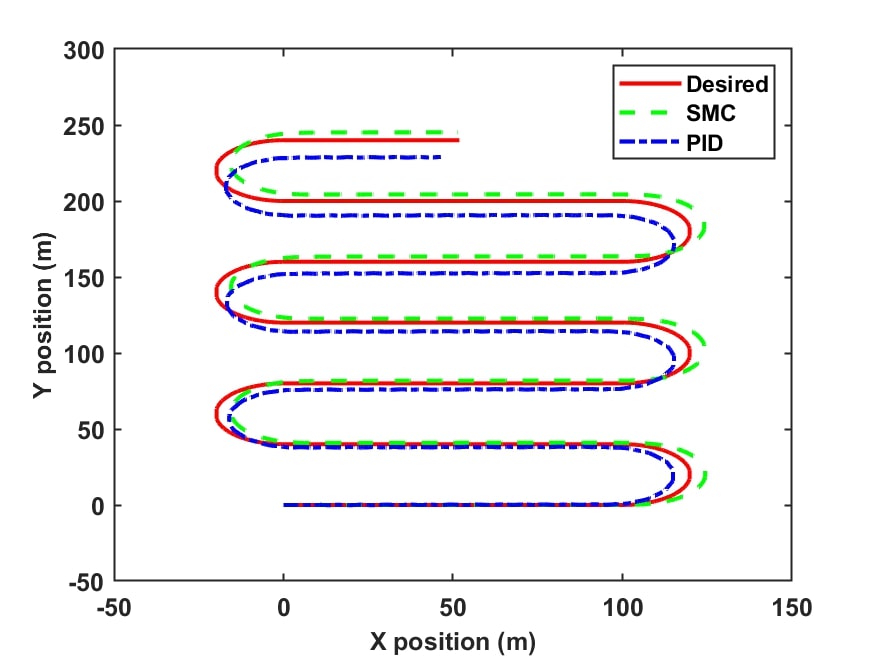} 
    \caption{Lawn mowing in $XY$ plane of SMC and Conventional controller with parameter variations}
    \label{fig_smcpid}
\end{figure}

Further, the comparison of the two strategies is shown in Figure \ref{fig_smcpid}.  The simulations encompass a simultaneous variation in two parameters, including variations in the Center of Gravity (CG) location ($x_{cg} = 8mm$) and a 50\% alteration in the axial drag coefficient $X_{u|u|}$. Additionally, a consistent external disturbance is introduced in the form of a constant ocean current, characterized by a relative velocity of $U_c = 0.1~m/s$, an angle of attack $\alpha_c = 0^\circ$, and a side-slip angle $\beta_c = 20^\circ$. Figure \ref{fig_smcpid} shows the paths followed in the $XY$ plane and highlights the difference between the desired path (shown in solid red) and the trajectories tracked with the conventional controller and SMC. Evidently, there's a noticeable deviation with the conventional controller, especially from the second cycle starting at $100 m$ on the Y-axis. On the other hand, the trajectory controlled by the SMC strategy remains remarkably steady, with minimal deviations from the intended path. These results strongly emphasize the robustness of the Sliding Mode Control (SMC) approach. It counteracts challenges posed by ocean currents, parameter variations, and shifts in the Center of Gravity (CG). The RMSE values are found to be 3.694 and 1.632 for conventional and SMC controllers respectively. It is evident that error with conventional controller has increased significantly compared to SMC. In conclusion, the SMC approach emerges as a reliable controller even in complex and changing underwater conditions.

\section{Conclusions} \label{scon}

In conclusion, this book chapter undertook an in-depth exploration of the modelling and robust control of an Autonomous Underwater Vehicle amidst the presence of uncertainties. The dynamic behaviour of the vehicle was represented with a 6-DoF model, accounting for the impact of ocean currents. The level flight requirements are derived across various speeds, and eventually 6-DoF model was dissected into horizontal and vertical subsystems, facilitating comprehensive linear analysis.

This study investigated different linear controllers, including depth, yaw, and speed subsystems. Simulation results were showcased, particularly highlighting the efficacy of these controllers in varying operational contexts, including level flight at a speed of 4 knots. Moreover, nonlinear control methodologies, encompassing both conventional and sliding-mode control were scrutinized. The controller design process diligently addressed a spectrum of uncertainties, ranging from ocean currents to parameter fluctuations, Center of Gravity (CG) deviations, and buoyancy variations. Importantly, a comparative assessment between the outcomes of conventional and SMC controllers was presented with a lawn-mowing manoeuvre scenario. This revealed SMC's enhanced robustness against disturbances and parameter uncertainties, albeit with the added complexity inherent to its design and implementation.

In essence, this book chapter delivered an encompassing study, elucidating the intricate interplay between modelling, control strategies, uncertainties, and robustness in the realm of AUVs. By delving into both linear and nonlinear approaches, the chapter provided valuable insights into the trade-offs between robustness and complexity, contributing to a holistic understanding of AUV control and navigation in challenging underwater environments.

\section*{Appendix} 
The myring hull parameters, force and moment coefficients of the research vehicle \cite{Pres2001}, \cite{HA2011} considered in the study are given in Table \ref{tpara} and Table \ref{tfc} respectively.

\begin{table}
\renewcommand{\arraystretch}{1.25}
\setlength{\tabcolsep}{7pt} 
\caption{Myring Parameters for AUV}
\centering
\begin{tabular}{|c|c|c|c|c|}
\hline
Description  & Unit       & Value       \\ \hline
Nose Length  ($a$) & $ m$     & 0.191       \\ \hline
Nose Offset ($a_{offset}$) &$ m$& 0.0165      \\ \hline
Mid-section Length ($b$) & $ m$ & 0.654       \\ \hline
Tail section Length  ($c$) & $ m$ & 0.541       \\ \hline
Tail Offset   ($c_{offset}$) &$ m$& 0.0368      \\ \hline
Exponential Coefficient ($n$) & - & 2           \\ \hline
Included angle at tail ($\theta$) & $ rad$ & 0.436       \\ \hline
Max Hull Diameter ($d$) & $ m$ & 0.191       \\ \hline
Vehicle Forward Length ($l_f$) & $ m$ & 0.828       \\ \hline
Total Vehicle Length  ($l$) & $ m$& 1.33        \\ \hline
\end{tabular} \label{tpara}
\end{table}

\begin{table}
\renewcommand{\arraystretch}{1.252}
\setlength{\tabcolsep}{5pt} 
\caption{Force and Moment coefficients of research vehicle}
\centering
\begin{tabular}{|c|c|c|c|c|c|}
\hline
Force Parameter   & Unit    & Value & Moment Parameter   & Unit    & Value      \\ \hline
$X_{uu}$    & $kg/m$    & -1.62  &$K_{pp}$           & $kg.m^2/rad^2$ & -0.001     \\ \hline
$X_{\dot{u}}$  & $kg$  & -0.93  &$K_{\dot{p}}$      & $kg.m^2/rad$  & -0.014    \\ \hline
$X_{wq}$  & $kg/rad$  & -35.5  &  $M_{ww}$           & $kg $            & 3.18   \\ \hline
$X_{qq}$  & $kg .m/rad$ & 1.93 &  $M_{qq}$  & $kg.m^2/rad^2$& -9.4                   \\ \hline
$X_{vr}$  & $kg/rad$  & 35.5   &   $M_{\dot{w}}$      & $kg.m$          & -1.93              \\ \hline
$X_{rr}$   & $kg .m/rad$          & -1.93  &$M_{uw}$           & $kg$             & 24         \\ \hline
$Y_{vv}$  & $kg/m$             & -131  & $M_{\dot{q}}$      & $kg.m^2/rad$  & -4.88    \\ \hline
$Y_{rr}$ & $kg .m/rad^2$  & 0.632  &$M_{uq}$           & $kg.m/rad$      & -2          \\ \hline
$Y_{uv}$ & $kg/m$          & -28.6   &$M_{vp}$           & $kg.m / rad$    & -1.93             \\ \hline
$Y_{\dot{v}}$ & $kg$ & -35.5  &  $M_{rp}$           & $kg.m^2/rad^2$ & 4.86   \\ \hline
$Y_{\dot{r}}$& $kg .m/rad$       & 1.93 &$M_{{uu}{\delta_s}}$ & $kg/rad $    & -6.15       \\ \hline
$Y_{ur}$ & $kg/rad$ & 5.22    &$N_{vv}$           & $kg$           & -3.18     \\ \hline
$Y_{wp}$ & $kg/rad$ & 35.5     &$N_{rr}$           & $kg.m^2/rad^2$ & -9.4    \\ \hline
$Y_{pq}$ & $kg .m/rad$    & 1.93 & $N_{uv}$           & $kg$             & -24     \\ \hline
$Y_{{uu}{\delta_r}}$ & $kg/(m .rad)$   & 9.64 & $N_{\dot{v}}$      & $kg.m$          & 1.93      \\ \hline
$Z_{ww}$ & $kg/m$        & -131 &$N_{\dot{r}}$       & $kg.m^2/rad^2$  & -4.88     \\ \hline
$Z_{qq}$ & $kg .m/rad^2$     & -0.632  &$N_{ur}$           & $kg.m / rad$     & -2          \\ \hline
$Z_{uw}$ & $kg/m$            & -28.6 &$N_{wp}$           & $kg.m / rad$     & -1.93            \\ \hline
$Z_{\dot{w}}$ & $kg $            & -35.5   &$N_{pq}$           & $kg.m^2/rad^2$  & -4.86    \\ \hline
$Z_{\dot{q}}$ & $kg .m/rad$       & -1.93 &$N_{{uu}{\delta_r}}$ & $kg/rad$        & -6.15       \\ \hline
$Z_{uq}$ & $kg/rad$    & -5.22   & - &- &-     \\ \hline
$Z_{vp}$ & $kg/rad$ & -35.5    & - &- &-   \\ \hline
$Z_{rp}$ & $kg/rad$  & 1.93   & - &- &-    \\ \hline
$Z_{{uu}{\delta_s}}$& $kg/(m.rad)$ & -9.64  & - &- &-     \\ \hline
\end{tabular} \label{tfc}
\end{table}


\begin{thebibliography}{10}

\bibitem{Von2003}
C.~von Alt.
\newblock Autonomous underwater vehicles.
\newblock {\em In: Autonomous Underwater Lagrangian Platforms and Sensors
  Workshop}, 3:2, 2003.

\bibitem{Widditsch1973}
H.~Widditsch.
\newblock {\em SPURV, The first decade}.
\newblock Washington Univ., 1973.

\bibitem{ASA1997}
B.~Allen, R.~Stokey, T.~Austin, N.~Forrester, R.~Goldsborough, M.~Purcell, and
  C.~von Alt.
\newblock Remus: a small, low cost auv; system description, field trials and
  performance results.
\newblock In {\em Oceans '97. MTS/IEEE Conference Proceedings}, volume~2, pages
  994--1000, 1997.

\bibitem{MBV2005}
M.~A. Moline, S.~M. Blackwell, C.~von Alt, B.~Allen, T.~Austin, J.~Case,
  N.~Forrester, R.~Goldsborough, M.~Purcell, and R.~Stokey.
\newblock Remote environmental monitoring units: An autonomous vehicle for
  characterizing coastal environments.
\newblock {\em Journal of Atmospheric and Oceanic Technology},
  22(11):1797--1808, 2005.

\bibitem{JLG2019}
F.~Jaffre, R.~Littlefield, M.~Grund, and M.~Purcell.
\newblock Development of a new version of the remus 6000 autonomous underwater
  vehicle.
\newblock In {\em OCEANS 2019 - Marseille}, pages 1--7, 2019.

\bibitem{McPhail2009}
S.~McPhail.
\newblock Autosub 6000: A deep diving long range auv.
\newblock {\em Journal of Bionic Engineering}, 6(1):55--62, 2009.

\bibitem{BPS2004}
E.A. {de Barros}, A.~Pascoal, and E.~{de Sa}.
\newblock Auv dynamics: Modelling and parameter estimation using analytical,
  semi-empirical, and cfd methods.
\newblock {\em IFAC Proceedings Volumes}, 37(10):369--376, 2004.

\bibitem{DMM2007}
E.~Desa, R.~Madhan, P~Maurya, G~Navelkar, AAMQ Mascarenhas, S~Prabhudesai,
  S~Afzulpurkar, and S.~N. Bandodkar.
\newblock The small maya auv--initial field results.
\newblock {\em International Ocean Systems}, 11(1), 2007.

\bibitem{SNPD2012}
S.~N. Shome, S.~Nandy, Pal D, S.~K. Das, S.~R.~K. Vadali, J.~Basu, and
  S.~Ghosh.
\newblock Development of modular shallow water auv: Issues \& trial results.
\newblock {\em Journal of The Institution of Engineers (India): Series C},
  93:217--228, 2012.

\bibitem{SMJ2012}
A.~Sousa, L.~Madureira, J.~Coelho, J.~Pinto, J.~Pereira, J.~B. Sousa, and
  P.~Dias.
\newblock {LAUV}: The man-portable autonomous underwater vehicle.
\newblock {\em IFAC Proceedings Volumes}, 45(5):268--274, 2012.

\bibitem{CLV2005}
J.~Curcio, J.~Leonard, J.~Vaganay, A.~Patrikalakis, A.~Bahr, D.~Battle,
  H.~Schmidt, and M.~Grund.
\newblock Experiments in moving baseline navigation using autonomous surface
  craft.
\newblock In {\em Proceedings of OCEANS 2005 MTS/IEEE}, pages 730--735. IEEE,
  2005.

\bibitem{Bluefin}
Naval Technology.
\newblock Bluefin 21 {AUV}.
\newblock
  https://www.naval-technology.com/projects/bluefin-21-autonomous-underwater-vehicle-auv/,
  \text{ Date last accessed: 30.07.2023}.

\bibitem{BluefinGD}
General Dynamics.
\newblock Bluefin 21 {AUV}.
\newblock
  https://gdmissionsystems.com/articles/2022/01/19/video-bluefin-21-uuv-navigates-autonomously-under-the-arctic-circle
  \text{Date last accessed: 30.07.2023}.

\bibitem{FGPB2005}
L.~E. Freitag, M.~Grund, J.~Partan, K.~Ball, S.~Singh, and P.~Koski.
\newblock Multi-band acoustic modem for the communications and navigation aid
  auv.
\newblock In {\em Proceedings of OCEANS 2005 MTS/IEEE}, pages 1080--1085, 2005.

\bibitem{Hugin}
Konsberg.
\newblock Hugin {AUV}.
\newblock
  https://www.naval-technology.com/projects/bluefin-21-autonomous-underwater-vehicle-auv/,
  \text{ Date last accessed: 30.07.2023}.

\bibitem{NP1990}
K.~S. Narendra and K.~Parthasarathy.
\newblock Identification and control of dynamical systems using neural
  networks.
\newblock {\em IEEE Transactions on Neural Networks}, 1(1):4--27, 1990.

\bibitem{KC1997}
B.~S. Kim and A.~J. Calise.
\newblock Nonlinear flight control using neural networks.
\newblock {\em Journal of Guidance, Control, and Dynamics}, 20(1):26--33, 1997.

\bibitem{SOM2005}
S.~Suresh, S.~N. Omkar, V.~Mani, and N.~Sundararajan.
\newblock Nonlinear adaptive neural controller for unstable aircraft.
\newblock {\em Journal of Guidance, Control, and Dynamics}, 28(6):1103--1111,
  2005.

\bibitem{KSO2009}
M.~V. Kumar, S.~Suresh, S.~N. Omkar, R.~Ganguli, and P.~Sampath.
\newblock A direct adaptive neural command controller design for an unstable
  helicopter.
\newblock {\em Engineering Applications of Artificial Intelligence},
  22(2):181--191, 2009.

\bibitem{KOG2006}
M.~V. Kumar, S.~N. Omkar, R.~Ganguli, P.~Sampath, and S.~Suresh.
\newblock Identification of helicopter dynamics using recurrent neural networks
  and flight data.
\newblock {\em Journal of the American Helicopter Society}, 51(2):164--174,
  2006.

\bibitem{RC1999}
R.~T. Rysdyk and A.~J. Calise.
\newblock Adaptive model inversion flight control for tilt-rotor aircraft.
\newblock {\em Journal of guidance, control, and dynamics}, 22(3):402--407,
  1999.

\bibitem{KCI2005}
A.~T. Kutay, A.~J. Calise, M.~Idan, and N.~Hovakimyan.
\newblock Experimental results on adaptive output feedback control using a
  laboratory model helicopter.
\newblock {\em IEEE Transactions on Control Systems Technology},
  13(2):196--202, 2005.

\bibitem{SK2008}
S.~Suresh and N.~Kannan.
\newblock Direct adaptive neural flight control system for an unstable unmanned
  aircraft.
\newblock {\em Applied Soft Computing}, 8(2):937--948, 2008.

\bibitem{DSS2022}
S.~Debarshi, S.~Sundaram, and N.~Sundararajan.
\newblock Robust emran-aided coupled controller for autonomous vehicles.
\newblock {\em Engineering Applications of Artificial Intelligence},
  110:104717, 2022.

\bibitem{CK2022}
N.~Cohen and I.~Klein.
\newblock Beamsnet: A data-driven approach enhancing doppler velocity log
  measurements for autonomous underwater vehicle navigation.
\newblock {\em Engineering Applications of Artificial Intelligence},
  114:105216, 2022.

\bibitem{KBR2020}
O.~K. Kinaci, I.~Bayezit, and M.~Reyhanoglu.
\newblock A practical feedforward speed control system for autonomous
  underwater vehicles.
\newblock {\em IEEE Journal of Ocean Engineering}, 218:108214, 2020.

\bibitem{Hydrus}
Advanced Navigation.
\newblock Hydrus.
\newblock https://www.advancednavigation.com/robotics/micro-auv/hydrus/,
  \text{Date last accessed : 21.08.2023}.

\bibitem{SSF2017}
Yang Shi, Chao Shen, Huazhen Fang, and Huiping Li.
\newblock Advanced control in marine mechatronic systems: A survey.
\newblock {\em IEEE/ASME Transactions on Mechatronics}, 22(3):1121--1131, 2017.

\bibitem{SDS2019}
A.~Sahoo, S.~K. Dwivedy, and P.~S. Robi.
\newblock Advancements in the field of autonomous underwater vehicle.
\newblock {\em IEEE Journal of Ocean Engineering}, 181:145--160, 2019.

\bibitem{KY2021}
H.~R. Karimi and Y.~Lu.
\newblock Guidance and control methodologies for marine vehicles: A survey.
\newblock {\em Control Engineering Practice}, 111:104785, 2021.

\bibitem{TCC2022}
A.~S. Tijjani, A.~Chemori, and V.~Creuze.
\newblock A survey on tracking control of unmanned underwater vehicles:
  Experiments-based approach.
\newblock {\em Annual Reviews in Control}, 54:125--147, 2022.

\bibitem{AXJ2023}
F.~Ahmed, X.~Xiang, C.~Jiang, G.~Xiang, and S.~Yang.
\newblock Survey on traditional and ai based estimation techniques for
  hydrodynamic coefficients of autonomous underwater vehicle.
\newblock {\em IEEE Journal of Ocean Engineering}, 268:113300, 2023.

\bibitem{Brunner1988}
G.~M. Brunner.
\newblock {\em Experimental verification of AUV performance}.
\newblock PhD thesis, Monterey, California. Naval Postgraduate School, 1988.

\bibitem{HG1992}
A.~J. Healey and M.~R. Good.
\newblock The nps auvii autonomous underwater vehicle testbed: Design and
  experimental verification.
\newblock {\em Naval Engineers Journal}, 104(3):191--202, 1992.

\bibitem{HM1992}
A.~J. Healey and D.~B. Marco.
\newblock Experimental verification of mission planning by autonomous mission
  execution and data visualization using the nps auv ii.
\newblock In {\em Proceedings of the 1992 Symposium on Autonomous Underwater
  Vehicle Technology}, pages 65--72, 1992.

\bibitem{Pres2001}
T.~Prestero.
\newblock Verification of a six-degree of freedom simulation model for the
  remus underwater vehicle.
\newblock Master's thesis, Massachusetts Institute of Technology, Cambridge,
  Massachusetts, USA, 2001.

\bibitem{herman2009}
P.~Herman.
\newblock Decoupled pd set-point controller for underwater vehicles.
\newblock {\em IEEE Journal of Ocean Engineering}, 36(6-7):529--534, 2009.

\bibitem{KHS2011}
F.~Sammut M.~Kokegei, M.~He.
\newblock Fully coupled 6 degree-of-freedom control of an over-actuated
  autonomous underwater vehicle.
\newblock {\em Autonomous underwater vehicles}, pages 147--170, 2011.

\bibitem{SDR2018}
Avilash Sahoo, S.~K. Dwivedy, and P.~S. Robi.
\newblock Dynamic modelling and control of a compact autonomous underwater
  vehicle.
\newblock In Anwar P.~P. Abdul~Majeed, Jessnor~Arif Mat-Jizat, Mohd Hasnun~Arif
  Hassan, Zahari Taha, Han~Lim Choi, and Junmo Kim, editors, {\em RITA 2018},
  pages 303--321, Singapore, 2020. Springer Singapore.

\bibitem{KB2015}
Mohammad~Hedayati Khodayari and Saeed Balochian.
\newblock Modeling and control of autonomous underwater vehicle (auv) in
  heading and depth attitude via self-adaptive fuzzy {PID} controller.
\newblock {\em Journal of Marine Science and Technology}, 20(3):559--578, 2015.

\bibitem{KFB2018}
E.~Kim, S.~Fan, and N.~Bose.
\newblock Autonomous underwater vehicle model-based high-gain observer for
  ocean current estimation.
\newblock In {\em 2018 IEEE/OES Autonomous Underwater Vehicle Workshop (AUV)},
  pages 1--6, 2018.

\bibitem{HGS2016}
R.~Hernandez-Alvarado, L.~G. Garcia-Valdovinos, T.~Salgado-Jimenez,
  A.~Gómez-Espinosa, and F.~F. Navarro.
\newblock Self-tuned {PID} control based on backpropagation neural networks for
  underwater vehicles.
\newblock In {\em OCEANS 2016 MTS/IEEE Monterey}, pages 1--5, 2016.

\bibitem{LZP2022}
Lu~Liu, Lichuan Zhang, Guang Pan, and Shuo Zhang.
\newblock Robust yaw control of autonomous underwater vehicle based on
  fractional-order {PID} controller.
\newblock {\em IEEE Journal of Ocean Engineering}, 257:111493, 2022.

\bibitem{YS1984}
D.~R.Yoerger and J.~J.Slotine.
\newblock Nonlinear trajectory control of autonomous underwater vehicles using
  the sliding methodology.
\newblock In {\em OCEANS 1984}, pages 588--593. IEEE, 1984.

\bibitem{HSC2010}
E.~Y. Hong, G.~H. Soon, and M.~Chitre.
\newblock Depth control of an autonomous underwater vehicle, starfish.
\newblock In {\em IEEE OCEANS - SYDNEY}, pages 1--6. IEEE, 2010.

\bibitem{JKY2014}
H.~Joe, M.~Kim, and S.-C. Yu.
\newblock Second-order sliding-mode controller for autonomous underwater
  vehicle in the presence of unknown disturbances.
\newblock {\em Nonlinear Dynamics}, 78:183--196, 2014.

\bibitem{TWTP2018}
Kantapon Tanakitkorn, Philip~A Wilson, Stephen~R Turnock, and Alexander~B
  Phillips.
\newblock Sliding mode heading control of an overactuated, hover-capable
  autonomous underwater vehicle with experimental verification.
\newblock {\em Journal of Field Robotics}, 35(3):396--415, 2018.

\bibitem{CPH1990}
F.~Papoulias R.~Cristi and A.~J. Healey.
\newblock Adaptive sliding mode control of autonomous underwater vehicles in
  the dive plane.
\newblock {\em IEEE Journal of Oceanic Engineering}, 15(3):152--160, 1990.

\bibitem{FF1995}
T.~I. Fossen and O.~Fjellstad.
\newblock Robust adaptive control of underwater vehicles: A comparative study.
\newblock {\em IFAC Proceedings Volumes}, 28(2):66--74, 1995.

\bibitem{YYCM2013}
Y.~Yang, K.~Yang, C.~Chen, L.~Mu, Y.~Chiu, C.~Yu, and W.~Yang.
\newblock Robust trajectory control for an autonomous underwater vehicle.
\newblock In {\em MTS/IEEE OCEANS - Bergen}, pages 1--9, 2013.

\bibitem{EZ2016}
M.~Zribi T.~Elmokadem and K.~Youcef-Toumi.
\newblock Trajectory tracking sliding mode control of underactuated auvs.
\newblock {\em Nonlinear Dynamics}, 84:1079--1091, 2016.

\bibitem{CZC2016}
Rongxin Cui, Xin Zhang, and Dong Cui.
\newblock Adaptive sliding-mode attitude control for autonomous underwater
  vehicles with input nonlinearities.
\newblock {\em IEEE Journal of Ocean Engineering}, 123:45--54, 2016.

\bibitem{GTC2019}
J.~Guerrero, J.~Torres, V.~Creuze, and A.~Chemori.
\newblock Trajectory tracking for autonomous underwater vehicle: An adaptive
  approach.
\newblock {\em IEEE Journal of Ocean Engineering}, 172:511--522, 2019.

\bibitem{TTA2021}
Pham~Nguyen {Nhut Thanh}, Phan~Minh Tam, and Ho~Pham {Huy Anh}.
\newblock A new approach for three-dimensional trajectory tracking control of
  under-actuated auvs with model uncertainties.
\newblock {\em IEEE Journal of Ocean Engineering}, 228:108951, 2021.

\bibitem{DYW2022}
P.~Du, W.~Yang, Y.~Wang, R.~Hu, Y.~Chen, and S.~H. Huang.
\newblock A novel adaptive backstepping sliding mode control for a lightweight
  autonomous underwater vehicle with input saturation.
\newblock {\em IEEE Journal of Ocean Engineering}, 263:112362, 2022.

\bibitem{Fossen2011}
T.~I. Fossen.
\newblock {\em Handbook of marine craft hydrodynamics and motion control}.
\newblock John Wiley \& Sons, 2011.

\bibitem{HA2011}
R.~Hall and S.~Anstee.
\newblock Trim calculation methods for a dynamical model of the remus 100
  autonomous underwater vehicle.
\newblock Technical report, Defence of Australia, 2011.

\bibitem{HRCMB1994}
A.~J. Healey, S.~M. Rock, S.~Cody, D.~Miles, and J.~P. Brown.
\newblock Toward an improved understanding of thruster dynamics for underwater
  vehicles.
\newblock In {\em Proceedings of IEEE Symposium on Autonomous Underwater
  Vehicle Technology (AUV'94)}, pages 340--352, 1994.

\end{thebibliography}

\end{document}